\documentclass[preprint,prd,aps,showpacs,showkeys,nofootinbib]{revtex4}
\usepackage{graphicx}
\usepackage{dcolumn}
\usepackage{bm}
\usepackage{soul}

\usepackage{color}
\usepackage[dvipsnames]{xcolor}
\usepackage{amssymb}

\definecolor{black-blue}{RGB}{77,116,175}
\definecolor{black-yellow}{RGB}{231,162,33}
\definecolor{black-green}{RGB}{144,180,58}
\definecolor{black-red}{RGB}{246,95,50}

\begin{document}

\title{Study on muon MDM and lepton EDM in BLMSSM via the mass insertion approximation}
\author{Xi Wang$^{1,2,3}$\footnote{wx$\_$0806@163.com}, Shu-Min Zhao$^{1,2,3}$\footnote{zhaosm@hbu.edu.cn}, Xin-Xin Long$^{1,2,3}$,
\\Yi-Tong Wang$^{1,2,3}$, Tong-Tong Wang$^{1,2,3}$, Hai-Bin Zhang$^{1,2,3}$, Tai-Fu Feng$^{1,2,3,4}$, Rong-Xiang Zhang$^{1,2,3}$\footnote{zrx@hbu.edu.cn}}

\affiliation{$^1$ Department of Physics, Hebei University, Baoding, 071002, China}
\affiliation{$^2$ Hebei Key Laboratory of High-precision Computation and Application of Quantum Field Theory, Baoding, 071002, China}
\affiliation{$^3$ Research Center for Computational Physics of Hebei Province, Baoding, 071002, China}
\affiliation{$^4$ Department of Physics, Chongqing University, Chongqing 401331, China}
\date{\today}

\begin{abstract}
In the framework of the MSSM extension with local gauged baryon and lepton numbers (BLMSSM), we calculate the muon anomalous magnetic dipole moment (MDM) and lepton $(e, \mu, \tau)$ electric dipole moment (EDM), and discuss how the muon MDM and lepton EDM depend on the parameters within the mass insertion approximation. Among many parameters, $\tan{\beta}$,~$g_L$,~$m_L$ and $\mu_H$ are more sensitive parameters for $a^{BL}_{\mu}$. Considering the experimental limitations, our best numerical result of $a^{BL}_{\mu}$ is around $2.5 \times 10^{-9}$, which can well compensate the departure between the experiment data and Standard Model (SM) prediction. The CP violating phases in BLMSSM are more than those in the MSSM, including new parameters $\theta_{\mu_L}$ and $\theta_{L}$. They can give large contributions, which play an important role in exploring the source of CP violation and probing new physics beyond SM.

\end{abstract}


\maketitle
\setcounter{page}{0}
\thispagestyle{empty}

\tableofcontents
\setcounter{page}{0}
\thispagestyle{empty}

\newpage

\section{Introduction}

In the development of the Standard Model (SM), the muon anomalous magnetic dipole moment (MDM) is an urgent problem to be solved, which indicates that there must be new physics beyond SM. The muon MDM is denoted by $a_\mu\equiv(g_\mu-2)/2$. The SM contributions to muon MDM have the following parts : 1.~the QED loop contributions \cite{g2rep2020,GWB,AKDN1,GCMH,MHBL,MDAH,AKDN2,TBPA,TAMH,GCFH,GECS,TBNC,TATK,ACWJ,CGDS};  2.~the electroweak contributions \cite{ACWJ,CGDS}; 3.~the hadronic vacuum polarization contributions  \cite{g2rep2020,GCMH, had2}; 4.~the hadronic light-by-light contributions \cite{GCFH, GECS, TBNC}. The specific expressions are as follows:
\begin{eqnarray}
&&a_\mu^{QED}=116584718.931(104)\times 10^{-11},
\nonumber\\&&a^{EW}_\mu=153.6(1.0)\times 10^{-11},
\nonumber\\&&a^{HVP}_\mu=6845(40)\times 10^{-11},
\nonumber\\&&a^{HLBL}_\mu=92(18)\times 10^{-11}.
\end{eqnarray}
Based on the above, SM prediction of muon anomaly is $a^{SM}_\mu=116591810(43)\times 10^{-11}$(0.37ppm) \cite{ g2rep2020, muon2, mdm2, TBPA}. New result on the muon MDM is reported by the E989 collaboration at Fermilab \cite{046}: $a_{\mu}^{FNAL}=116592040(54)\times 10^{-11}$(0.46ppm) and 3.3 standard deviations larger than the SM prediction, which is in great agreement with BNL E821 result \cite{GWB}. The new averaged experiment value of
 muon anomaly is $a^{exp}_{\mu}=116592061(41)\times 10^{-11}$(0.35ppm).
Combining all available measurements, we now obtain 4.2$\sigma$ deviation between the experiment and SM expectation ($\Delta a_\mu=a^{exp}_\mu-a^{SM}_\mu=251(59)\times 10^{-11}$).

The study of lepton electric dipole moment (EDM) is used to probe the source of CP violation \cite{EDM1}. The latest experiment shows that the upper bound of electron EDM is $|d^{exp}_e|$ $<$ $1.1 \times 10^{-29}$ e.cm at the $90 \%$ confidence level \cite{de,de1,de2}, the muon EDM is $|d^{exp}_{\mu}|$ $<$ $1.8 \times 10^{-19}$ e.cm at the $95 \%$ confidence level and the tau EDM is $|d^{exp}_{\tau}|$ $<$ $1.1 \times 10^{-17}$ e.cm at the $95 \%$ confidence level \cite{pdg2022}. The estimated SM value for the electron EDM is about $|d_{e}|\simeq10^{-38}$ e.cm \cite{deSM1,deSM2}, which
is too small to be detected by the current experiments. Therefore, if large electron EDM is
probed, one can ensure it is the sign of new physics beyond SM.

The study of muon MDM and lepton EDM has very important practical significance for exploring new physics. Some studies investigate the supersymmetric (SUSY) one-loop contributions to muon MDM. The authors \cite{susy1, susy2} obtain the
approximate SUSY one-loop contributions by simplification
\begin{eqnarray}
|a_{\mu}^{SUSY}|=13\times10^{-10}\Big(
\frac{100{\rm GeV}} {M_{SUSY}}\Big)^2\tan\beta\texttt{sign}[\mu_H]
\label{susyoneloop}.
\end{eqnarray}
Here, $M_{SUSY}$ represents the masses of neutralinos, charginos and scalar leptons of the second generation, which are equal. The SUSY contributions can be easily evaluated from Eq.~(\ref{susyoneloop}). The authors \cite{wx3} study the muon g-2 in
several GUT-scale constrained SUSY models, such as CMSSM/mSUGRA, pMSSM, CMSSM/mSUGRA extensions and GMSB/AMSB extensions. The numerical
result of muon g-2 is researched with the MultiNest technique for the parameter space \cite{caojj,caojj1,caojj2} in the GNMSSM with a singlino-dominated neutralino as a dark matter candidate. The muon g-2 is further studied under $\mathbb{Z}_3$-NMSSM with LHC analyses in Ref.~\cite{caojj3}. They study to what extent the g-2 can be explained in anomaly mediation scenarios \cite{YW1}. Even
if there is no new particle in this energy range, one can measure the g-2 directly via the channel to a Higgs boson and a monochromatic photon \cite{YW2}. Next, let's briefly review our previous work on muon MDM. We study the corrections from loop diagrams to muon MDM with the mass eigenstate in the BLMSSM and B-LMSSM \cite{muon,o2,o3}. With the effective Lagrangian method \cite{slh, 04,o1}, we also research the contributions to muon MDM from loop diagrams under the $U(1)_X$SSM and $\mu \nu$SSM.

To better explain the CP violation mechanism \cite{NPdl1,NPdl2,NPdl3,NPdl4}, scientists are trying to find CP violating terms in new physics beyond SM. In the MSSM \cite{mssm,mssm1,mssm2,Z2015}, there are several CP violating phases, which can contribute significantly to the lepton EDM. The normal size CP violating phases $O(1)$ and particle mass in the TeV range can cause the electron EDM to exceed the current experimental upper limit ($|d^{exp}_e| < 1.1\times10^{-29}$ e.cm). In order to rectify this situation, there are three ways: the first is to make the phases small $O(10^{-2}-10^{-3})$, the second is to increase the particle mass to the several 10 TeV range, and the third is to make internal cancellations between phases \cite{EDM4,EDM5}.

In the extension of SM, the MSSM \cite{mssm} is one of the most widely studied models. The authors propose the extension of the MSSM  with local gauged B and L (BLMSSM) \cite{BLMSSM00, BLMSSM000}, where the baryon and lepton
numbers are local gauge symmetries spontaneously broken at the TeV scale. The BLMSSM has two advantages: one is that the broken baryon number (B) can explain asymmetry of matter-antimatter in the universe, and the other is that the broken lepton number (L) can generate tiny neutrino mass by the seesaw mechanism. In BLMSSM, the proton decay can be avoided by discrete symmetry called matter parity and R-parity \cite{BLMSSM11}.

In this paper, we investigate the BLMSSM contributions to the
muon MDM and lepton $(e, \mu, \tau)$ EDM via the mass insertion approximation (MIA). In the process of analysis, we show the mass eigenstate expressions of muon MDM, the MIA expressions of muon MDM and lepton EDM in detail. We discuss the numerical difference between the mass eigenstate expressions and the MIA expressions to prove the accuracy of the latter. In comparison, the MIA makes it easier and more intuitive to observe sensitive parameters. However, in the BLMSSM, the one-loop corrections are similar to the MSSM results in analytic form. The difference is that the BLMSSM contributions have the new gaugino $\lambda_{L}$ and gauge coupling constant $g_L$. Under the latest  experimental constraints, our results can well enough compensate the deviation of muon MDM and satisfy the experimental limitations of lepton EDM.

The rest of the paper is organized as follows. In Section II, we briefly summarize the main components of the BLMSSM. In Section III, we show analytic forms of the BLMSSM contributions to the muon MDM ($a_{\mu}^{BL}$) and the lepton EDM ($d_{l}^{BL}$). In Section IV, some
numerical results are shown. The last section is devoted to summary.

\section{The BLMSSM}

The local gauge group of BLMSSM is $SU(3)_{C}\otimes SU(2)_{L}\otimes U(1)_{Y}\otimes U(1)_{B}\otimes U(1)_{L}$ \cite{BLMSSM1,BLMSSM2}. Compared with MSSM, BLMSSM includes exotic quarks $(\hat{Q}_{4},\hat{U}_{4}^c,\hat{D}_{4}^c,\hat{Q}_{5}^c,\hat{U}_{5},\hat{D}_{5})$ and exotic leptons $(\hat{L}_{4},\hat{E}_{4}^c,\hat{N}_{4}^c,\hat{L}_{5}^c,\hat{E}_{5},\hat{N}_{5})$, which are used to eliminate B and L anomaly, respectively. The exotic Higgs ($\hat{\Phi}_{B},\hat{\varphi}_{B}$) are used to break $B$ spontaneously with nonzero vacuum expectation values (VEVs), and the exotic Higgs ($\hat{\Phi}_{L},\hat{\varphi}_{L}$) are used to break $L$ spontaneously with nonzero VEVs. The superfields $\hat{X}$ and $\hat{X}^\prime$ are used to make the exotic quarks unstable. The right-handed neutrinos $N_R^c$ are introduced to provide tiny masses of neutrinos through the seesaw mechanism. Table \ref {quarks} displays these additional fields in detail.

 \begin{table}[ht]
\caption{ The new fields in the BLMSSM.}
\begin{center}
\begin{tabular}{|c|c|c|c|c|c|}
\hline
Superfields & $SU(3)_C$ & $SU(2)_L$ & $U(1)_Y$ & $U(1)_B$ & $U(1)_L$\\
\hline
\hline
$\hat{Q}_4$ & 3 & 2 & 1/6 & $B_4$ & 0 \\
\hline
$\hat{U}^c_4$ & $\bar{3}$ & 1 & -2/3 & -$B_4$ & 0 \\
\hline
$\hat{D}^c_4$ & $\bar{3}$ & 1 & 1/3 & -$B_4$ & 0 \\
\hline
$\hat{Q}_5^c$ & $\bar{3}$ & 2 & -1/6 & -$(1+B_4)$ & 0 \\
\hline
$\hat{U}_5$ & $3$ & 1 & 2/3 &  $1 + B_4$ & 0 \\
\hline
$\hat{D}_5$ & $3$ & 1 & -1/3 & $1 + B_4$ & 0 \\
\hline
$\hat{L}_4$ & 1 & 2 & -1/2 & 0 & $L_4$ \\
\hline
$\hat{E}^c_4$ & 1 & 1 & 1 & 0 & -$L_4$ \\
\hline
$\hat{N}^c_4$ & 1 & 1 & 0 & 0 & -$L_4$ \\
\hline
$\hat{L}_5^c$ & 1 & 2 & 1/2 & 0 & -$(3 + L_4)$ \\
\hline
$\hat{E}_5$ & 1 & 1 & -1 & 0 & $3 + L_4$ \\
\hline
$\hat{N}_5$ & 1 & 1 & 0 & 0 & $3 + L_4$ \\
\hline
$\hat{\Phi}_B$ & 1 & 1 & 0 & 1 & 0 \\
\hline
$\hat{\varphi}_B$ & 1 & 1 & 0 & -1 & 0 \\
\hline
$\hat{\Phi}_L$ & 1 & 1 & 0 & 0 & -2 \\
\hline
$\hat{\varphi}_L$ & 1 & 1 & 0 & 0 & 2 \\
\hline
$\hat{X}$ & 1 & 1 & 0 & $2/3 + B_4$ & 0 \\
\hline
$\hat{X'}$ & 1 & 1 & 0 & -$(2/3 + B_4)$ & 0 \\
\hline
$\hat{N}_R^c$ & 1 & 1 & 0 & 0 & -1 \\
\hline
\end{tabular}
\end{center}
\label{quarks}
\end{table}

In the BLMSSM, the superpotential is expressed as \cite{weBLMSSM}
\begin{eqnarray}
&&{\cal W}_{{BLMSSM}}={\cal W}_{{MSSM}}+{\cal W}_{B}+{\cal W}_{L}+{\cal W}_{X}\;,
\label{superpotential1}
\end{eqnarray}
where, ${\cal W}_{{MSSM}}$ is the superpotential of the MSSM. The concrete forms of ${\cal W}_{B},{\cal W}_{L},{\cal W}_{X}$ are
\begin{eqnarray}
&&{\cal W}_{B}=\lambda_{Q}\hat{Q}_{4}\hat{Q}_{5}^c\hat{\Phi}_{B}+\lambda_{U}\hat{U}_{4}^c\hat{U}_{5}
\hat{\varphi}_{B}+\lambda_{D}\hat{D}_{4}^c\hat{D}_{5}\hat{\varphi}_{B}+\mu_{B}\hat{\Phi}_{B}\hat{\varphi}_{B}
\nonumber\\
&&\hspace{1.2cm}
+Y_{{u_4}}\hat{Q}_{4}\hat{H}_{u}\hat{U}_{4}^c+Y_{{d_4}}\hat{Q}_{4}\hat{H}_{d}\hat{D}_{4}^c
+Y_{{u_5}}\hat{Q}_{5}^c\hat{H}_{d}\hat{U}_{5}+Y_{{d_5}}\hat{Q}_{5}^c\hat{H}_{u}\hat{D}_{5}\;,
\nonumber\\
&&{\cal W}_{L}=Y_{{e_4}}\hat{L}_{4}\hat{H}_{d}\hat{E}_{4}^c+Y_{{\nu_4}}\hat{L}_{4}\hat{H}_{u}\hat{N}_{4}^c
+Y_{{e_5}}\hat{L}_{5}^c\hat{H}_{u}\hat{E}_{5}+Y_{{\nu_5}}\hat{L}_{5}^c\hat{H}_{d}\hat{N}_{5}
\nonumber\\
&&\hspace{1.2cm}
+Y_{\nu}\hat{L}\hat{H}_{u}\hat{N}^c+\lambda_{{N^c}}\hat{N}^c\hat{N}^c\hat{\varphi}_{L}
+\mu_{L}\hat{\Phi}_{L}\hat{\varphi}_{L}\;,
\nonumber\\
&&{\cal W}_{X}=\lambda_1\hat{Q}\hat{Q}_{5}^c\hat{X}+\lambda_2\hat{U}^c\hat{U}_{5}\hat{X}^\prime
+\lambda_3\hat{D}^c\hat{D}_{5}\hat{X}^\prime+\mu_{X}\hat{X}\hat{X}^\prime\;.
\label{superpotential-BL}
\end{eqnarray}

The local gauge symmetry $SU(3)_{C}\otimes SU(2)_{L}\otimes U(1)_{Y}\otimes U(1)_{B}\otimes U(1)_{L}$ can break down to the electromagnetic symmetry $U(1)_{e}$, when the $SU(2)_L$ doublets ($H_{u}$, $H_{d}$) and singlets ($\Phi_{B}$, $\varphi_{B}$, $\Phi_{L}$, $\varphi_{L}$) obtain nonzero VEVs $\upsilon_{u},\;\upsilon_{d}$
and $\upsilon_{{B}},\;\overline{\upsilon}_{{B}},\;\upsilon_{L},\;\overline{\upsilon}_{L}$ respectively. The $SU(2)_L$ doublets and singlets are shown as
\begin{eqnarray}
&&H_{u}=\left(\begin{array}{c}H_{u}^+\\{1\over\sqrt{2}}\Big(\upsilon_{u}+H_{u}^0+iP_{u}^0\Big)\end{array}\right)\;,~~~~
H_{d}=\left(\begin{array}{c}{1\over\sqrt{2}}\Big(\upsilon_{d}+H_{d}^0+iP_{d}^0\Big)\\H_{d}^-\end{array}\right)\;,
\nonumber\\&&\Phi_{B}={1\over\sqrt{2}}\Big(\upsilon_{B}+\Phi_{B}^0+iP_{B}^0\Big)\;,~~~~~~~~~
\varphi_{B}={1\over\sqrt{2}}\Big(\overline{\upsilon}_{B}+\varphi_{B}^0+i\overline{P}_{B}^0\Big)\;,
\nonumber\\&&\Phi_{L}={1\over\sqrt{2}}\Big(\upsilon_{L}+\Phi_{L}^0+iP_{L}^0\Big)\;,~~~~~~~~~~
\varphi_{L}={1\over\sqrt{2}}\Big(\overline{\upsilon}_{L}+\varphi_{L}^0+i\overline{P}_{L}^0\Big)\;.
\label{VEVs}
\end{eqnarray}

The soft breaking terms of BLMSSM are shown as follows \cite{BLMSSM1, BLMSSM2, BLMSSM000}
\begin{eqnarray}
&&{\cal L}_{{soft}}={\cal L}_{{soft}}^{MSSM}-(m_{{\tilde{\nu}^c}}^2)_{{IJ}}\tilde{N}_I^{c*}\tilde{N}_J^c
-m_{{\tilde{Q}_4}}^2\tilde{Q}_{4}^\dagger\tilde{Q}_{4}-m_{{\tilde{U}_4}}^2\tilde{U}_{4}^{c*}\tilde{U}_{4}^c
-m_{{\tilde{D}_4}}^2\tilde{D}_{4}^{c*}\tilde{D}_{4}^c
\nonumber\\
&&\hspace{1.3cm}
-m_{{\tilde{Q}_5}}^2\tilde{Q}_{5}^{c\dagger}\tilde{Q}_{5}^c-m_{{\tilde{U}_5}}^2\tilde{U}_{5}^*\tilde{U}_{5}
-m_{{\tilde{D}_5}}^2\tilde{D}_{5}^*\tilde{D}_{5}-m_{{\tilde{L}_4}}^2\tilde{L}_{4}^\dagger\tilde{L}_{4}
-m_{{\tilde{\nu}_4}}^2\tilde{N}_{4}^{c*}\tilde{N}_{4}^c
\nonumber\\
&&\hspace{1.3cm}
-m_{{\tilde{e}_4}}^2\tilde{E}_{_4}^{c*}\tilde{E}_{4}^c-m_{{\tilde{L}_5}}^2\tilde{L}_{5}^{c\dagger}\tilde{L}_{5}^c
-m_{{\tilde{\nu}_5}}^2\tilde{N}_{5}^*\tilde{N}_{5}-m_{{\tilde{e}_5}}^2\tilde{E}_{5}^*\tilde{E}_{5}
-m_{{\Phi_{B}}}^2\Phi_{B}^*\Phi_{B}
\nonumber\\
&&\hspace{1.3cm}
-m_{{\varphi_{B}}}^2\varphi_{B}^*\varphi_{B}-m_{{\Phi_{L}}}^2\Phi_{L}^*\Phi_{L}
-m_{{\varphi_{L}}}^2\varphi_{L}^*\varphi_{L}-\Big(m_{B}\lambda_{B}\lambda_{B}
+m_{L}\lambda_{L}\lambda_{L}+h.c.\Big)
\nonumber\\
&&\hspace{1.3cm}
+\Big\{A_{{u_4}}Y_{{u_4}}\tilde{Q}_{4}H_{u}\tilde{U}_{4}^c+A_{{d_4}}Y_{{d_4}}\tilde{Q}_{4}H_{d}\tilde{D}_{4}^c
+A_{{u_5}}Y_{{u_5}}\tilde{Q}_{5}^cH_{d}\tilde{U}_{5}+A_{{d_5}}Y_{{d_5}}\tilde{Q}_{5}^cH_{u}\tilde{D}_{5}
\nonumber\\
&&\hspace{1.3cm}
+A_{{BQ}}\lambda_{Q}\tilde{Q}_{4}\tilde{Q}_{5}^c\Phi_{B}+A_{{BU}}\lambda_{U}\tilde{U}_{4}^c\tilde{U}_{5}\varphi_{B}
+A_{{BD}}\lambda_{D}\tilde{D}_{4}^c\tilde{D}_{5}\varphi_{B}+B_{B}\mu_{B}\Phi_{B}\varphi_{B}
+h.c.\Big\}
\nonumber\\
&&\hspace{1.3cm}
+\Big\{A_{{e_4}}Y_{{e_4}}\tilde{L}_{4}H_{d}\tilde{E}_{4}^c+A_{{\nu_4}}Y_{{\nu_4}}\tilde{L}_{4}H_{u}\tilde{N}_{4}^c
+A_{{e_5}}Y_{{e_5}}\tilde{L}_{5}^cH_{u}\tilde{E}_{5}+A_{{\nu_5}}Y_{{\nu_5}}\tilde{L}_{5}^cH_{d}\tilde{N}_{5}
\nonumber\\
&&\hspace{1.3cm}
+A_{N}Y_{\nu}\tilde{L}H_{u}\tilde{N}^c+A_{{N^c}}\lambda_{{N^c}}\tilde{N}^c\tilde{N}^c\varphi_{L}
+B_{L}\mu_{L}\Phi_{L}\varphi_{L}+h.c.\Big\}
\nonumber\\
&&\hspace{1.3cm}
+\Big\{A_1\lambda_1\tilde{Q}\tilde{Q}_{5}^cX+A_2\lambda_2\tilde{U}^c\tilde{U}_{5}X^\prime
+A_3\lambda_3\tilde{D}^c\tilde{D}_{5}X^\prime+B_{X}\mu_{X}XX^\prime+h.c.\Big\}\;.
\label{soft-breaking}
\end{eqnarray}

The used mass matrices can be found in the works \cite{Z2015,FM}.
The relevant Feynman rules same as those in MSSM for the present computation are collected in Ref. \cite{FM}. The Feynman rules for vertices uniquely used in the BLMSSM are as follows:
\begin{figure}[ht]
\setlength{\unitlength}{5.0mm}
\centering
\includegraphics[width=4.8in]{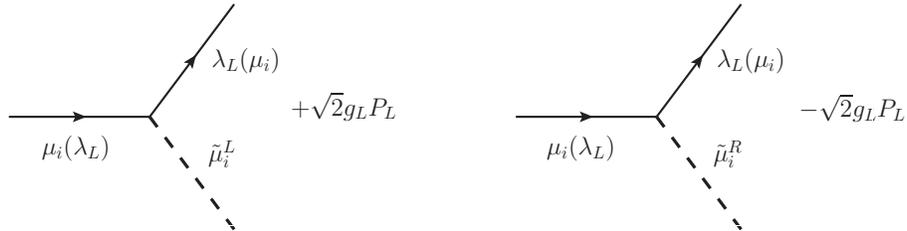}
\caption{Feynman rules for the unique vertices in the BLMSSM.}\label{N0}
\end{figure}

\section{Formulation}

\subsection{The muon MDM}
The lepton MDM can be obtained from the following effective Lagrangian by using the on-shell condition for the external leptons,
\begin{eqnarray}
&&{\cal L}_{{MDM}}={e\over4m_{l}}\;a_{l}\;\bar{l}\sigma^{\mu\nu}
l\;F_{{\mu\nu}}\label{adm},
\end{eqnarray}
with $\sigma^{\mu\nu}=i[{\gamma}_{\mu},{\gamma}_{\nu}]/2$. $e$ and $l$ denote the electric charge and the lepton fermion, respectively. $F_{{\mu\nu}}$ is the electromagnetic field strength, and $m_{l}$ is the lepton mass.

The Feynman amplitude can be expressed by these dimension 6 operators \cite{lepton} with the effective Lagrangian method for the process $l^I\rightarrow l^I+\gamma$. The dimension 8 operators are suppressed by additional factor $\frac{m_{l}^2}{M_{SUSY}^2}$ $\sim$ ($10^{-7}$, $10^{-8}$) , which are neglected safely. Therefore, these dimension 6 operators are enough to use in future calculations. The operators related to lepton MDM are $\mathcal{O}_{2,3,6}^{L,R}$. The lepton MDM is the combination of the Wilson coefficients $C^{L,R}_{2,3,6}$. Here, $\mathcal{D}_{\mu}=\partial_{\mu}+ieA_{\mu}$ and $P_{L,R}=\frac{1\mp\gamma_5}{2}$. The specific forms of those dimension 6 operators are
\begin{eqnarray}
&&\mathcal{O}_1^{L,R}=\frac{1}{(4\pi)^2}\bar{l}(i\mathcal{D}\!\!\!\slash)^3P_{L,R}l,~~~~~~~~~~~~~~
\mathcal{O}_2^{L,R}=\frac{eQ_f}{(4\pi)^2}\overline{(i\mathcal{D}_{\mu}l)}\gamma^{\mu}
F\cdot\sigma P_{L,R}l,
\nonumber\\
&&\mathcal{O}_3^{L,R}=\frac{eQ_f}{(4\pi)^2}\bar{l}F\cdot\sigma\gamma^{\mu}
P_{L,R} (i\mathcal{D}_{\mu}l),~~~~\mathcal{O}_4^{L,R}=\frac{eQ_f}{(4\pi)^2}\bar{l}(\partial^{\mu}F_{\mu\nu})\gamma^{\nu}
P_{L,R}l,\nonumber\\&&
\mathcal{O}_5^{L,R}=\frac{m_l}{(4\pi)^2}\bar{l}(i\mathcal{D}\!\!\!\slash)^2P_{L,R}l,
~~~~~~~~~~~~~~\mathcal{O}_6^{L,R}=\frac{eQ_fm_l}{(4\pi)^2}\bar{l}F\cdot\sigma
P_{L,R}l.\label{operators}
\end{eqnarray}

\subsubsection{The mass eigenstate}

\begin{figure}[ht]
\setlength{\unitlength}{5.0mm}
\centering
\includegraphics[width=4in]{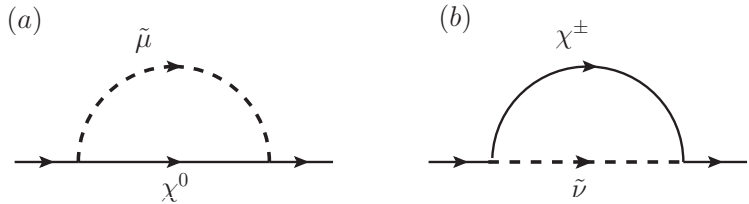}
\caption{ The two one-loop diagrams drawn in the mass eigenstate for $l^I\rightarrow l^I+\gamma$. The external photon line has to be attached to the charged internal lines.
}\label{N1}
\end{figure}

The analytical forms of one-loop corrections in BLMSSM are similar to that of MSSM. The differences are: 1.~the squared mass matrixes of scalar leptons because of
new parameters $g_L,\bar{v}_L,v_L$ and so on; 2.~three-generation right-handed neutrinos are introduced,  which lead to the neutrinos and scalar neutrinos are doubled. In the BLMSSM, there are four parts contributing to muon MDM: 1.~scalar muon ($\tilde{\mu}$) and neutralino (${\chi}^0$) [Fig. \ref{N1}(a)]; 2.~scalar neutrino ($\tilde{\nu}$) and chargino (${\chi}^{\pm}$) [Fig. \ref{N1}(b)]; 3.~neutral Higgs ($H^0$) and muon ($\mu$); 4.~charged Higgs ($H^{\pm}$) and neutrino ($\nu$).

The one-loop Higgs contribution to muon MDM is very small, because it is inhibited by the factor $\frac{m_{\mu}^2}{m^2_W}$. The mass matrix of neutrinos and the squared mass matrix of scalar neutrinos are extended to $6\times6$. From these analysis, the contributions of type 3 and 4 are entirely negligible. Due to the mass of the new vector boson $Z_L$ being greater than 5.1~TeV, the one-loop contributions from $Z_L$-muon are suppressed by the factor $\frac{m_Z^2}{m^2_{Z_L}}\sim 4\times 10^{-4}$. So, we neglect $Z_L$-muon one-loop contributions.

Therefore, the one-loop new physics contributions to muon MDM are given entirely by the Fig.~\ref{N1}. On the basis of the one-loop self-energy diagrams, we can get the one-loop triangle diagrams by attaching a photon on the internal line in all possible ways. These diagrams have been comprehensively discussed in the BLMSSM with the mass eigenstate \cite{muon}, and the exact results have been derived. We show the general results in the form:
\begin{eqnarray}
&&a_{\mu}^{BL}=a_{\mu}^{\tilde{\mu}\chi^{0}}+a_{\mu}^{\tilde{\nu}\chi^{\pm}},
\end{eqnarray}
with
\begin{eqnarray}
&&a_{\mu}^{\tilde{\mu}\chi^{0}}=
-\frac{e^2}{2s_W^2}\sum_{i=1}^6\sum_{j=1}^4\Big[\texttt{Re}[(\mathcal{S}_1)^I_{ij}(\mathcal{S}_2)^{I*}_{ij}]
\sqrt{x_{\chi_j^{0}}x_{l^I}}x_{\tilde{L}_i}\frac{\partial^2 \mathcal{B}(x_{\chi_j^{0}},x_{\tilde{L}_i})}{\partial x_{\tilde{L}_i}^2}
\nonumber\\&&\hspace{1.4cm}+\frac{1}{3}(|(\mathcal{S}_1)^I_{ij}|^2+|(\mathcal{S}_2)^I_{ij}|^2)x_{\tilde{L}_i}x_{l^I}
\frac{\partial\mathcal{B}_1(x_{\chi_j^{0}},x_{\tilde{L}_i})}{\partial x_{\tilde{L}_i}}\Big],
\end{eqnarray}
\begin{eqnarray}
&&a_{\mu}^{\tilde{\nu}\chi^{\pm}}=\frac{e^2}{s_W^2}\sum_{J=1}^3\sum_{i,j=1}^2
\Big[\sqrt{2}\frac{m_{l^I}}{m_W}\texttt{Re}[Z_+^{1j}Z_-^{2j}]|Z_{\tilde{\nu}^{IJ}}^{1i}|^2\sqrt{x_{\chi_j^{\pm}}x_{l^I}}
\mathcal{B}_1(x_{\tilde{\nu}^{Ji}},x_{\chi_j^{\pm}})
\nonumber\\&&\hspace{1.4cm}+\frac{1}{3}(|Z_+^{1j}Z_{\tilde{\nu}^{IJ}}^{1i*}|^2+\frac{m_{l^I}^2}{2m^2_W}|Z_-^{2j*}Z_{\tilde{\nu}^{IJ}}^{1i*}|^2)
x_{\chi_j^{\pm}}x_{l^I}\frac{\partial\mathcal{B}_1(x_{\tilde{\nu}^{Ji}},x_{\chi_j^{\pm}})}{\partial x_{\chi_j^{\pm}}}\Big].
\end{eqnarray}
Here, $x_i=\frac{m_i^2}{M_{SUSY}^2}$, $m_i$ is the particle mass. The abbreviation notations $s_W=\sin\theta_W,~c_W=\cos\theta_W$, where $\theta_W$ is the Weinberg angle. We define the functions $\mathcal{B}(x,y),\;\mathcal{B}_1(x,y)$
\begin{eqnarray}
\mathcal{B}(x,y)=\frac{1}{16 \pi
   ^2}\Big(\frac{x \ln x}{y-x}+\frac{y \ln
   y}{x-y}\Big),~~~
\mathcal{B}_1(x,y)=(
\frac{\partial }{\partial y}+\frac{y}{2}\frac{\partial^2 }{\partial y^2})\mathcal{B}(x,y).
\end{eqnarray}
The couplings $(\mathcal{S}_1)^I_{ij},\;(\mathcal{S}_2)^{I}_{ij}$ are shown as
\begin{eqnarray}
&&(\mathcal{S}_1)^I_{ij}=\frac{1}{c_W}Z_{\tilde{L}}^{Ii*}(Z_N^{1j}s_W+Z_N^{2j}c_W)-\frac{m_{l^I}}{\cos\beta m_W}Z_{\tilde{L}}^{(I+3)i*}Z_N^{3j},\nonumber\\&&
(\mathcal{S}_2)^I_{ij}=-2\frac{s_W}{c_W}Z_{\tilde{L}}^{(I+3)i*}Z_N^{1j*}-\frac{m_{l^I}}{\cos\beta m_W}Z_{\tilde{L}}^{Ii*}Z_N^{3j*}.
\end{eqnarray}
The matrices $Z_{\tilde{L}},~Z_N$ diagonalize the mass matrices of scalar lepton and neutralino, respectively. $Z_-,~Z_+$ are used to diagonalize the chargino mass matrix. The mass squared matrix of scalar neutrino are diagonalized by $Z_{\tilde{\nu}^{IJ}}$.

\subsubsection{The mass insertion approximation}

Through the above discussion of BLMSSM contributions to muon MDM, we can know that the contributions do not represent an enhancement proportional to $\frac{m_\chi}{m_{l^I}}$, because it is suppressed by the combined rotation matrixes. In fact, they produce an overall enhancement factor $\tan\beta$ \cite{dabeta1,dabeta2}. In other words, $|a_{\mu}^{BL}|$ becomes large as $\tan\beta$ increases. Thus, it's more convenient to use the mass insertion approximation (MIA)  \cite{FM,wx1,wx7,dabeta1} to calculate, and the role of parameters can be more clearly displayed. However, the mass eigenstate in the previous section is more appropriate for an exact evaluation. Now, we obtain the specific forms of the one-loop contributions by using MIA in the BLMSSM.

\begin{figure}[ht]
\setlength{\unitlength}{5.0mm}
\centering
\includegraphics[width=5in]{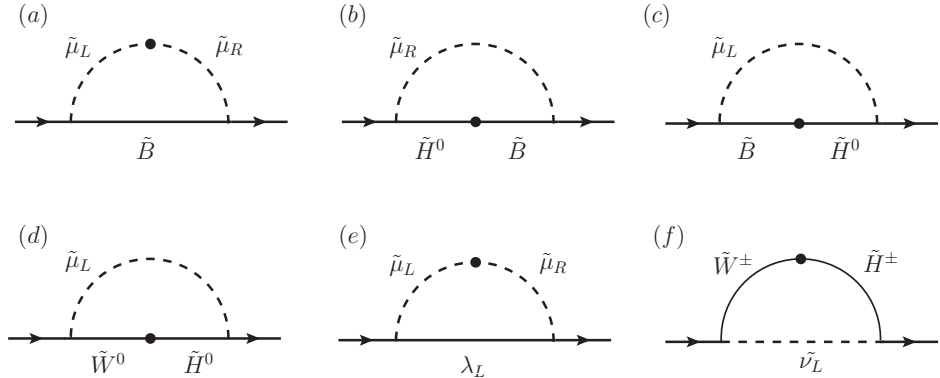}
\caption{ Feynman diagrams for generating muon MDM and lepton EDM based on MIA. The external photons are connected to the charged internal lines in all possible ways.}\label{N2}
\end{figure}

a. The one-loop contributions from $\tilde{B}$-$\tilde{\mu}_L$-$\tilde{\mu}_R$.
\begin{eqnarray}
&&a^{BL,(a)}_\mu=2 g_{1}^2 x_{\mu} \sqrt{ x_1 x_{\mu_{H}}} \tan{\beta}[I_1(x_1,x_{\tilde{\mu}_L},x_{\tilde{\mu}_R})+
I_2(x_1,x_{\tilde{\mu}_L},x_{\tilde{\mu}_R})\nonumber\\
&&\hspace{1.7cm}-J_1(x_1,x_{\tilde{\mu}_L},x_{\tilde{\mu}_R})
-J_2(x_1,x_{\tilde{\mu}_L},x_{\tilde{\mu}_R})-J_3(x_1,x_{\tilde{\mu}_L},x_{\tilde{\mu}_R})]\label{MIAa}.
\end{eqnarray}
with $m_1=m_{\tilde{B}}$. The functions $I_1(x,y,z)$, $I_2(x,y,z)$, $J_1(x,y,z)$, $J_2(x,y,z)$ and $J_3(x,y,z)$ are defined as
\begin{eqnarray}
&&I_1(x,y,z)=\frac{1}{16 \pi ^2} \Big\{\frac{ \left(z^2-x y\right)\log z}{(x-z)^2 (y-z)^2}-\frac{1}{(x-z)(y-z)}
\nonumber\\&&\hspace{2.3cm}-\frac{x \log x}{(x-y) (x-z)^2}+\frac{y \log y}{(x-y)(y-z)^2}\Big\},
\\&&I_2(x,y,z)=
  \frac{1} {16 \pi ^2} \Big\{\frac{1}{(x-y)
   (y-z)}-\frac{x \log x}{(x-y)^2 (x-z)}\nonumber\\&&\hspace{2.3cm}+\frac{ \left(y^2-x z\right)\log y}{(x-y)^2 (y-z)^2}+\frac{z \log z}{(x-z)
   (y-z)^2}\Big\},
\\&&J_1(x,y,z)=
   \frac{1}{32 \pi ^2} \Big\{\frac{x (z-3 y)+z (y+z)}{(x-z)^2 (y-z)^2}-\frac{2 x^2 \log x}{(x-y) (x-z)^3}\nonumber\\&&\hspace{2.4cm}+\frac{2 y^2 \log y}{(x-y)
   (y-z)^3}-\frac{2 [x^2 y^2+z^3 (x+y)-3 x y z^2]\log z}{(x-z)^3
   (y-z)^3}\Big\},
\\&&J_2(x,y,z)=
\frac{1}{32 \pi ^2} \Big\{\frac{x (y-3 z)+y (y+z)}{(x-y)^2 (y-z)^2}-\frac{2 x^2 \log x}{(x-y)^3 (x-z)}\nonumber\\&&\hspace{2.4cm}+\frac{2 [x^2 z^2+x y^2 (y-3 z)+y^3 z]\log y }{(x-y)^3
   (y-z)^3}+\frac{2 z^2 \log z}{(x-z)
   (z-y)^3}\Big\},
\\&&J_3(x,y,z)=
   \frac{1}{16 \pi ^2} \Big\{\frac{x (y+z)-2 y z}{(x-y)
   (x-z) (y-z)^2}-\frac{x^2 \log x}{(x-y)^2 (x-z)^2}\nonumber\\&&\hspace{2.4cm}+\frac{y [y (y+z)-2 x z]\log y}{(x-y)^2 (y-z)^3}+\frac{z [z (y+z)-2 x y] \log
   z}{(x-z)^2 (z-y)^3}\Big\}.
\end{eqnarray}

b. The one-loop contributions from $\tilde{B}$-$\tilde{H}^0$-$\tilde{\mu}_R$.
\begin{eqnarray}
&&a^{BL,(b)}_\mu=-2 g_{1}^2 x_{\mu} \sqrt{ x_1 x_{\mu_{H}}} \tan{\beta}[I_1(x_1,x_{{\mu}_H},x_{\tilde{\mu}_R})
-J_1(x_1,x_{{\mu}_H},x_{\tilde{\mu}_R})]\label{MIAb}.
\end{eqnarray}

c. The one-loop contributions from $\tilde{B}$-$\tilde{H}^0$-$\tilde{\mu}_L$.
\begin{eqnarray}
&&a^{BL,(c)}_\mu= g_{1}^2 x_{\mu} \sqrt{ x_1 x_{\mu_{H}}} \tan{\beta}[I_1(x_1,x_{{\mu}_H},x_{\tilde{\mu}_L})
-J_1(x_1,x_{{\mu}_H},x_{\tilde{\mu}_L})]\label{MIAc}.
\end{eqnarray}

d. The one-loop contributions from $\tilde{W}^0$-$\tilde{H}^0$-$\tilde{\mu}_L$.
\begin{eqnarray}
&&a^{BL,(d)}_\mu=- g_{2}^2 x_{\mu} \sqrt{ x_2 x_{\mu_{H}}} \tan{\beta}[I_1(x_2,x_{{\mu}_H},x_{\tilde{\mu}_L})
-J_1(x_2,x_{{\mu}_H},x_{\tilde{\mu}_L})]\label{MIAd}.
\end{eqnarray}
here, $m_2=m_{\tilde{W}^0}=m_{\tilde{W}^{\pm}}$.

e. The one-loop contributions from ${\lambda}_L$-$\tilde{\mu}_L$-$\tilde{\mu}_R$.
\begin{eqnarray}
&&a^{BL,(e)}_\mu=-4 g_{L}^2 x_{\mu} \sqrt{ x_L x_{\mu_{H}}} \tan{\beta}[I_1(x_L,x_{\tilde{\mu}_L},x_{\tilde{\mu}_R})+
I_2(x_L,x_{\tilde{\mu}_L},x_{\tilde{\mu}_R})\nonumber\\
&&\hspace{1.7cm}-J_1(x_L,x_{\tilde{\mu}_L},x_{\tilde{\mu}_R})
-J_2(x_L,x_{\tilde{\mu}_L},x_{\tilde{\mu}_R})-J_3(x_L,x_{\tilde{\mu}_L},x_{\tilde{\mu}_R})]\label{MIAe}.
\end{eqnarray}

f. The one-loop contributions from $\tilde{W}^{\pm}$-$\tilde{H}^{\pm}$-$\tilde{\nu}_L$.
\begin{eqnarray}
&&a^{BL,(f)}_\mu\!=2 g_{2}^2 x_{\mu} \sqrt{ x_2 x_{\mu_{H}}} \tan{\beta}[J_2(x_2,x_{{\mu}_H},x_{\tilde{\nu}_L})
\!+J_4(x_2,x_{{\mu}_H},x_{\tilde{\nu}_L})\!+J_5(x_2,x_{{\mu}_H},x_{\tilde{\nu}_L})]\label{MIAf}.
\end{eqnarray}
We define the functions $J_4(x,y,z)$ and $J_5(x,y,z)$ as
\begin{eqnarray}
&&J_4(x,y,z)=
\frac{1}{16 \pi ^2} \Big\{\frac{z (x+y)-2 x y}{(x-y)^2 (x-z) (y-z)}+\frac{x [x (x+y)-2 y z]\log x }{(x-y)^3 (x-z)^2}\nonumber\\
&&\hspace{2.3cm}+\frac{y [y (x+y)-2 x z]\log
   y}{(y-x)^3 (y-z)^2}-\frac{z^2 \log z}{(x-z)^2 (y-z)^2}\Big\},
\\&&J_5(x,y,z)=
\frac{1}{32 \pi ^2} \Big\{\frac{x^2+x (y+z)-3 y z}{(x-y)^2 (x-z)^2}-\frac{2
   [x^3 (y+z)-3 x^2 y z+y^2 z^2]\log x}{(x-y)^3 (x-z)^3}\nonumber\\
&&\hspace{2.3cm}+\frac{2 y^2 \log
   y}{(x-y)^3 (y-z)}+\frac{2 z^2 \log z}{(x-z)^3 (z-y)}\Big\}.
\end{eqnarray}

The one-loop contributions to muon MDM can be expressed as
\begin{eqnarray}
&&a_{\mu}^{\tilde{\mu}\chi^{0}} \simeq a^{BL,(a)}_\mu+a^{BL,(b)}_\mu+a^{BL,(c)}_\mu+a^{BL,(d)}_\mu+a^{BL,(e)}_\mu,
\nonumber\\&&a_{\mu}^{\tilde{\nu}\chi^{\pm}} \simeq a^{BL,(f)}_\mu.
\end{eqnarray}
We ought to notice that the contributions to muon MDM are related to $\tan\beta$ and $x_i$ in the Eqs.~(\ref{MIAa}),~(\ref{MIAb})$-$(\ref{MIAf}). This situation is consistent with MSSM. The contribution related to the new gaugino $\lambda_{L}$ is shown in Eq.~(\ref{MIAe}), which includes the new gauge coupling constant $g_L$. Furthermore, we obtain the conclusion that $a^{BL,(a)}_\mu$, $a^{BL,(e)}_\mu$ and $a^{BL,(f)}_\mu$ occupy the dominant position after numerical comparison. When $m_{{\lambda}_L}$ is negative, the signs of $a^{BL,(a)}_\mu$, $a^{BL,(e)}_\mu$ and $a^{BL,(f)}_\mu$ are the same. We can get the reasonable corrections of new physics.

\subsubsection{Degenerate result}

Next, we assume that all the masses of the superparticles are almost degenerate to more clearly know the influential factor on $a_{\mu}^{BL}$. The masses of superparticles ($m_1,~m_2,~\mu_H,~m_{\tilde{\nu}_L}
,~m_{\tilde{\mu}_R},~m_{\tilde{\mu}_L},~m_L$) are equal to $M_{SUSY}$ \cite{04}:
\begin{eqnarray}
&&|m_1|=|m_2|=|\mu_H|=m_{\tilde{\nu}_L}
=m_{\tilde{\mu}_R}=m_{\tilde{\mu}_L}=|m_L|=M_{SUSY}.
\end{eqnarray}
The functions can be simplified as
\begin{eqnarray}
&&I_1(1,1,1)=I_2(1,1,1)=\frac{1}{96\pi^2}, \nonumber\\&&J_1(1,1,1)=J_2(1,1,1)=J_3(1,1,1)=J_4(1,1,1)=J_5(1,1,1)=\frac{1}{192\pi^2}.
\end{eqnarray}

The one-loop MSSM results (chargino-sneutrino, neutralino-smuon) in this case are consistent with the results of Ref. \cite{dabeta1}. Here, $\texttt{sign}[m_1]=\texttt{sign}[m_2]=\texttt{sign}[\mu_H]=1$.
\begin{eqnarray}
a^{MSSM}_{\mu}\simeq\frac{1}{192\pi^2}\frac{m_\mu^2}{M_{SUSY}^2}(g_1^2+5g_2^2)\tan\beta.\label{amumssm1}
\end{eqnarray}
In the BLMSSM, the one-loop results of muon MDM are given by
\begin{eqnarray}
&&a^{BL}_{\mu}\simeq\frac{1}{192\pi^2}\frac{m_\mu^2}{M_{SUSY}^2}(g_1^2+5g_2^2)\tan\beta
\nonumber\\&&\hspace{1.3cm}-\frac{1}{48\pi^2}\frac{m_\mu^2}{M_{SUSY}^2} g_L^2 \tan\beta \texttt{sign}[\mu_H m_L ].\label{amumssm2}
\end{eqnarray}
The corrections can reach large value, when $\texttt{sign}[m_1]=\texttt{sign}[m_2]=\texttt{sign}[\mu_H]=1$ and $\texttt{sign}[m_L]=-1$.
\begin{eqnarray}
&&a^{BL}_{\mu}\rightarrow\frac{1}{192\pi^2}\frac{m_\mu^2}{M_{SUSY}^2}(g_1^2+5g_2^2+4g_L^2)\tan\beta.\label{amumssm3}
\end{eqnarray}

\begin{figure}[ht]
\setlength{\unitlength}{5mm}
\centering
\includegraphics[width=3.1in]{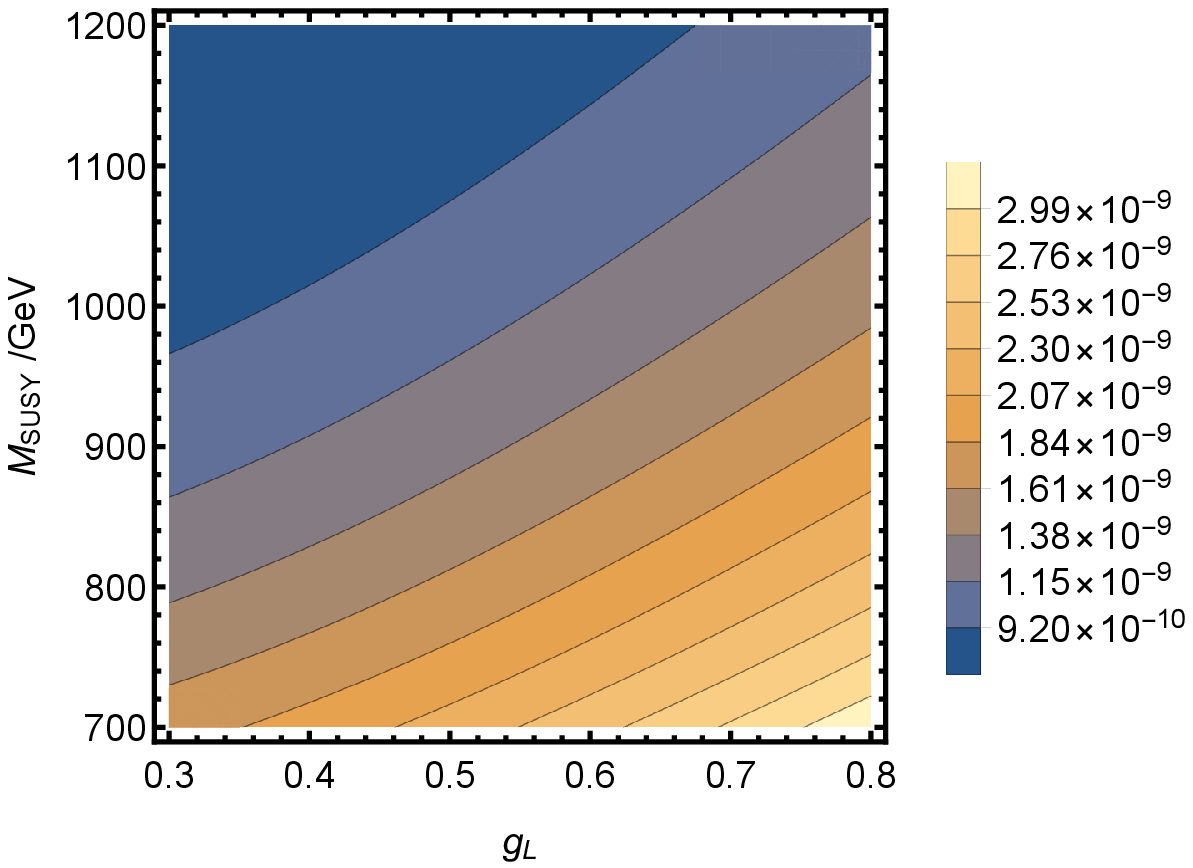}
\vspace{0.2cm}
\setlength{\unitlength}{5mm}
\centering
\includegraphics[width=3.1in]{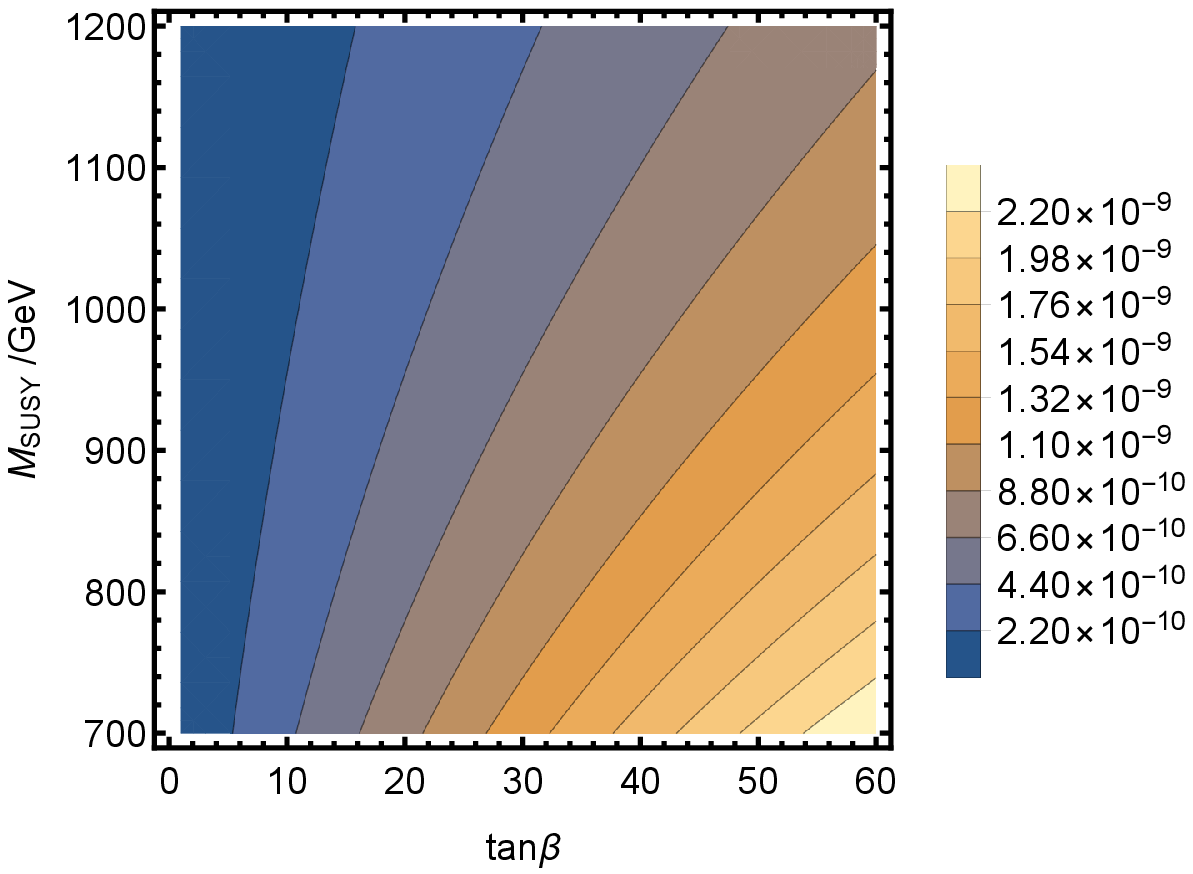}
\vspace{0.2cm}
\setlength{\unitlength}{5mm}
\centering
\includegraphics[width=3.1in]{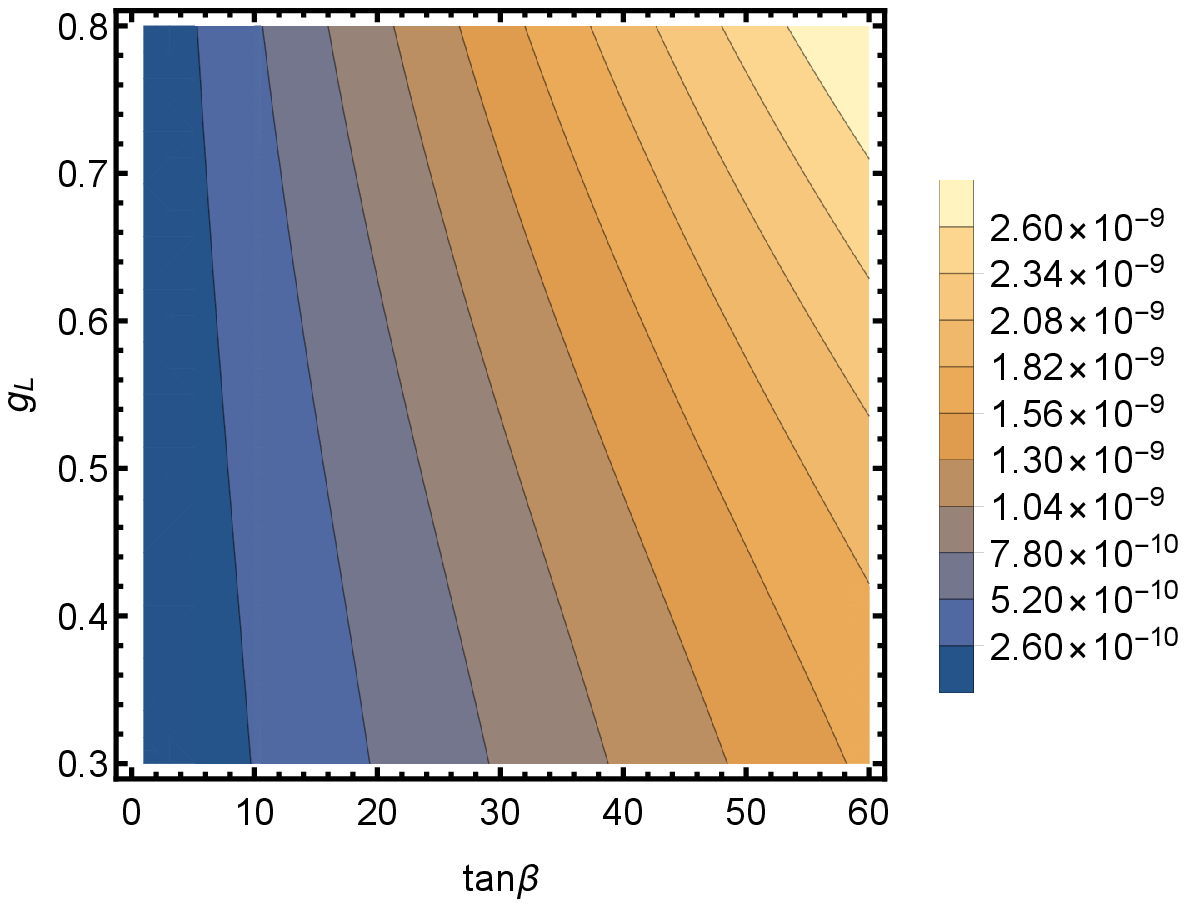}
\caption{The effects of $M_{SUSY}$, $\tan\beta$ and $g_L$ on $a^{BL}_{\mu}$. The upper left figure denotes $g_L$-$M_{SUSY}$ plane with $\tan\beta=50$. The upper right figure denotes $\tan\beta$-$M_{SUSY}$ plane with $g_L=0.45$. The bottom figure denotes $\tan\beta$-$g_L$ plane with $M_{SUSY}=1000~{\rm GeV}$.}{\label {to}}
\end{figure}

According to the above expressions, we study the effect of $M_{SUSY}$, $\tan\beta$ and $g_L$ on the BLMSSM contributions to muon MDM. The results are shown in Fig.~\ref {to}.
First, we plot the results for $\tan\beta=50$ in the $g_L$-$M_{SUSY}$ plane. As we can see, if we take a smaller value of $M_{SUSY}$, the $a^{BL}_{\mu}$ is enhanced in the large $g_L$ region. Next, the upper right figure denotes $\tan\beta$-$M_{SUSY}$ plane when $g_L=0.45$. The results imply that large $\tan\beta$ and small $M_{SUSY}$ can produce suitable BLMSSM corrections to compensate the departure. At last, the bottom figure shows the results in the  plane of $\tan\beta$ versus $g_L$. When the values of $\tan\beta$ and $g_L$ enlarge, the value of $a^{BL}_{\mu}$ also increases, but $\tan\beta$ is more sensitive than $g_L$. It shows that $M_{SUSY}$, $\tan\beta$ and $g_L$ are sensitive, and have a direct effect on $a^{BL}_{\mu}$.

\subsection{The lepton EDM}

The lepton EDM can be obtained by using effective Lagrange method, and the Feynman amplitudes can be expressed by these dimension 6 operators in Eq.~(\ref{operators}).
Adopting on-shell condition for external leptons, only
$\mathcal{O}_{2,3,6}^{\mp}$ contribute to lepton EDM. This is consistent with muon MDM. The lepton EDM is expressed as \begin{eqnarray}
&&{\cal L}_{{EDM}}=-{i\over2}d_{l}\overline{l}\sigma^{\mu\nu}\gamma_5
lF_{{\mu\nu}}
\label{eq1}.
\end{eqnarray}

In the BLMSSM, there are essentially two types of one-loop triangle diagrams which contribute to $d^{BL}_l$ : 1. the neutralino-slepton diagram; 2. the
chargino-sneutrino diagram. Using the MIA, we can obtain six diagrams that have major contributions to $d^{BL}_{l}$, which are shown in Fig.~\ref {N2}. These contributions can be given by
\begin{eqnarray}
&&d^{BL,(a)}_l= \frac{e g_{1}^2} {M_{SUSY}} \sqrt{x_{l} x_{1}  x_{\mu_{H}}} e^{i*\theta_1} e^{i*\theta_{\mu_{H}}} \tan{\beta}[I_1(x_1,x_{\tilde{L}_L},x_{\tilde{L}_R})+
I_2(x_1,x_{\tilde{L}_L},x_{\tilde{L}_R})\nonumber\\
&&\hspace{1.7cm}-J_1(x_1,x_{\tilde{L}_L},x_{\tilde{L}_R})
-J_2(x_1,x_{\tilde{L}_L},x_{\tilde{L}_R})-J_3(x_1,x_{\tilde{L}_L},x_{\tilde{L}_R})],
\nonumber\\&&d^{BL,(b)}_l= \frac{e g_{1}^2} {M_{SUSY}} \sqrt{x_{l} x_{1} x_{\mu_{H}}} e^{i*\theta_1} e^{i*\theta_{\mu_{H}}} \tan{\beta} [I_1(x_1,x_{{\mu}_H},x_{\tilde{L}_R})
-J_1(x_1,x_{{\mu}_H},x_{\tilde{L}_R})],
\nonumber\\&&d^{BL,(c)}_l= - \frac{e g_{1}^2} {2 M_{SUSY}} \sqrt{x_{l} x_{1}  x_{\mu_{H}}}  e^{i*\theta_1} e^{i*\theta_{\mu_{H}}} \tan{\beta}[I_1(x_1,x_{{\mu}_H},x_{\tilde{L}_L})
-J_1(x_1,x_{{\mu}_H},x_{\tilde{L}_L})],
\nonumber\\&&d^{BL,(d)}_l=  \frac{e g_{2}^2} {2 M_{SUSY}} \sqrt{x_{l} x_{2}  x_{\mu_{H}}}  e^{i*\theta_2} e^{i*\theta_{\mu_{H}}}  \tan{\beta}[I_1(x_2,x_{{\mu}_H},x_{\tilde{L}_L})
-J_1(x_2,x_{{\mu}_H},x_{\tilde{L}_L})],
\nonumber\\&&d^{BL,(e)}_l=-\frac{2 e g_{L}^2} {M_{SUSY}} \sqrt{x_{l} x_L  x_{\mu_{L}}}  e^{i*\theta_L} e^{i*\theta_{\mu_{L}}} \tan{\beta}[I_1(x_L,x_{\tilde{L}_L},x_{\tilde{L}_R})+
I_2(x_L,x_{\tilde{L}_L},x_{\tilde{L}_R})\nonumber\\
&&\hspace{1.7cm}-J_1(x_L,x_{\tilde{L}_L},x_{\tilde{L}_R})
-J_2(x_L,x_{\tilde{L}_L},x_{\tilde{L}_R})-J_3(x_L,x_{\tilde{L}_L},x_{\tilde{L}_R})],
\nonumber\\&&d^{BL,(f)}_l=- \frac{ e g_{2}^2} {M_{SUSY}} \sqrt{x_{l} x_{2} x_{\mu_{H}}}  e^{i*\theta_2} e^{i*\theta_{\mu_{H}}} \tan{\beta}[J_2(x_2,x_{{\mu}_H},x_{\tilde{\nu}_L})
+J_4(x_2,x_{{\mu}_H},x_{\tilde{\nu}_L})
\nonumber\\&&\hspace{1.7cm}+J_5(x_2,x_{{\mu}_H},x_{\tilde{\nu}_L})].\label{edmjiexi}
\end{eqnarray}
The one-loop contributions to lepton EDM can be expressed as
\begin{eqnarray}
&&d_{l}^{BL}=d_{l}^{\tilde{l}\chi^{0}}+d_{l}^{\tilde{\nu}\chi^{\pm}},
\nonumber\\&&d_{l}^{\tilde{l}\chi^{0}} \simeq d^{BL,(a)}_l+d^{BL,(b)}_l+d^{BL,(c)}_l+d^{BL,(d)}_l+d^{BL,(e)}_l,
\nonumber\\&&d_{l}^{\tilde{\nu}\chi^{\pm}} \simeq d^{BL,(f)}_l.
\end{eqnarray}

In Eq.~(\ref{edmjiexi}), some parameters are defined as follows
\begin{eqnarray}
&&m_{\tilde{B}}=M_1*e^{i*\theta_1},~m_{\tilde{W}^0}=m_{\tilde{W}^{\pm}}=M_2*e^{i*\theta_2},~\mu_H=M_{\mu_H}*e^{i*\theta_{\mu_H}},
\nonumber\\&&\mu_L=M_{\mu_L}*e^{i*\theta_{\mu_L}},~ m_L=M_L*e^{i*\theta_{L}},~x_l=\frac{m_l^2}{M_{SUSY}^2},
~x_{\tilde{\nu}_L}=\frac{m_{\tilde{\nu}_L}^2}{M_{SUSY}^2},
\nonumber\\&&x_{\tilde{L}_L}=\frac{m_{\tilde{L}_L}^2}{M_{SUSY}^2},
~~x_{\tilde{L}_R}=\frac{m_{\tilde{L}_R}^2}{M_{SUSY}^2},~~x_i=\frac{|M_i|^2}{M_{SUSY}^2}.
\end{eqnarray}
$M_i$ means the above five particle masses in the form of complex function ($M_1$, $M_2$, $M_{\mu_H}$, $M_{\mu_L}$, $M_L$). $\theta_1,\theta_2$ and $\theta_{\mu_H}$ are the CP violating phases of the parameters $m_{\tilde{B}},~m_{\tilde{W}^0}~(m_{\tilde{W}^{\pm}}$), and $\mu_H$. $\theta_{\mu_L}$ and $\theta_{L}$ are the CP violating phases of new parameters $\mu_L$ and $m_L$. In these formulas, the CP violating phases are conspicuous, and we can more easily observe the cancellation between the CP violating phases.

\section{Numerical results}

In this section, we first discuss the numerical difference between the  mass eigenstate and the mass insertion approximation. Then, the one-loop contributions of the muon MDM and lepton $(e, \mu, \tau)$ EDM are discussed numerically via the MIA. In the numerical discussion, we consider the latest experimental limitations of particles \cite{wx1,wx2,wx4,wx5,wx6}. The lightest CP-even Higgs mass $m_{h^0}=125.1~{\rm GeV}$ \cite{su1,su2}. The slepton mass is greater than $700~{\rm GeV}$, and the chargino mass is greater than $1100 ~{\rm GeV}$ \cite{wx7}. Taking $Z_L$ boson mass is greater than $5.1~{\rm TeV}$ to satisfy the mass constraint from LHC experiments \cite{Zp5d1}.

\subsection{The relative error between the two methods}

In order to determine the accuracy of the MIA expressions, we discuss the numerical difference between the mass eigenstate expressions and the MIA expressions from the point value and one-dimensional graph. Firstly, we discuss the point value. We set the same parameters of the two expressions to the same values, i.e., $\tan{\beta}=50$,~$m_1=300~{\rm GeV}$,~$m_2=1100~{\rm GeV}$ and $g_L=1/3$, and adjust the other parameters to make the masses of particles in the two methods meet the mass limits while maintaining roughly equal. The $a^{BL}_{\mu}$ obtained by the mass eigenstate expressions and the MIA expressions are $2.279\times10^{-9}$ and $2.257\times10^{-9}$, respectively. The relative error ($2.279\times10^{-9}-2.257\times10^{-9}\over2.279\times10^{-9}$) is about $0.96 \%$, which is relatively small.

\begin{figure}[ht]
\setlength{\unitlength}{5mm}
\centering
\includegraphics[width=3.03in]{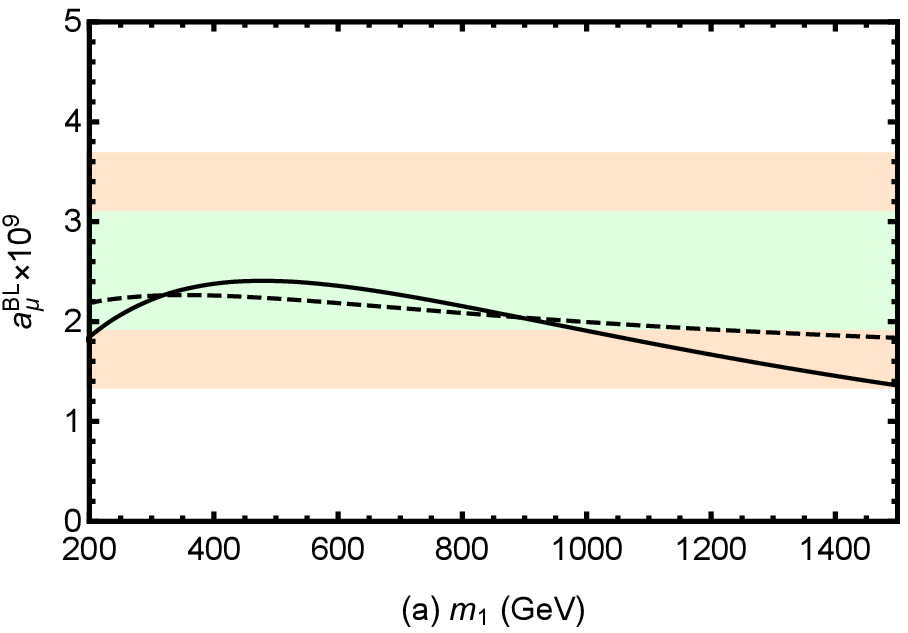}
\vspace{0.2cm}
\setlength{\unitlength}{5mm}
\centering
\includegraphics[width=3.1in]{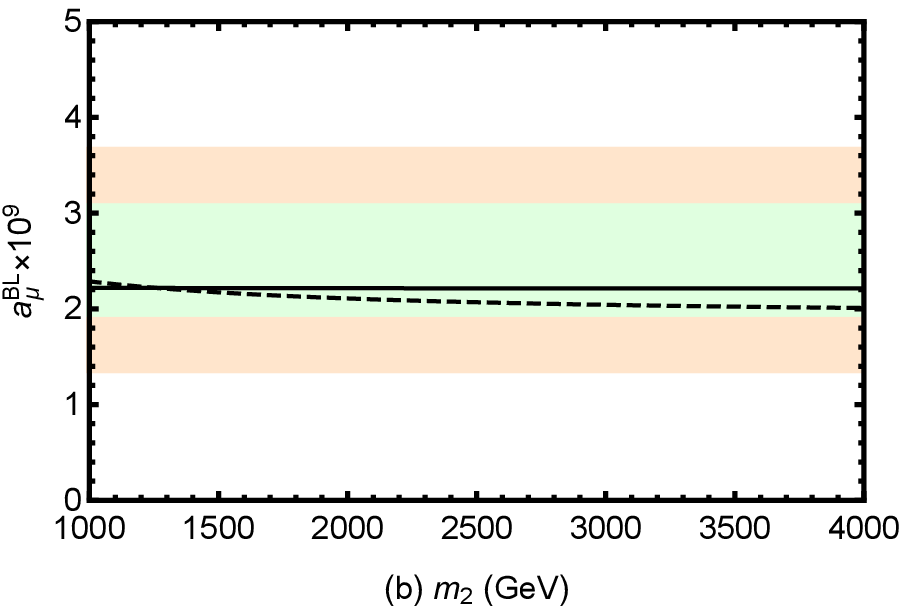}
\vspace{0.2cm}
\setlength{\unitlength}{5mm}
\centering
\includegraphics[width=3.1in]{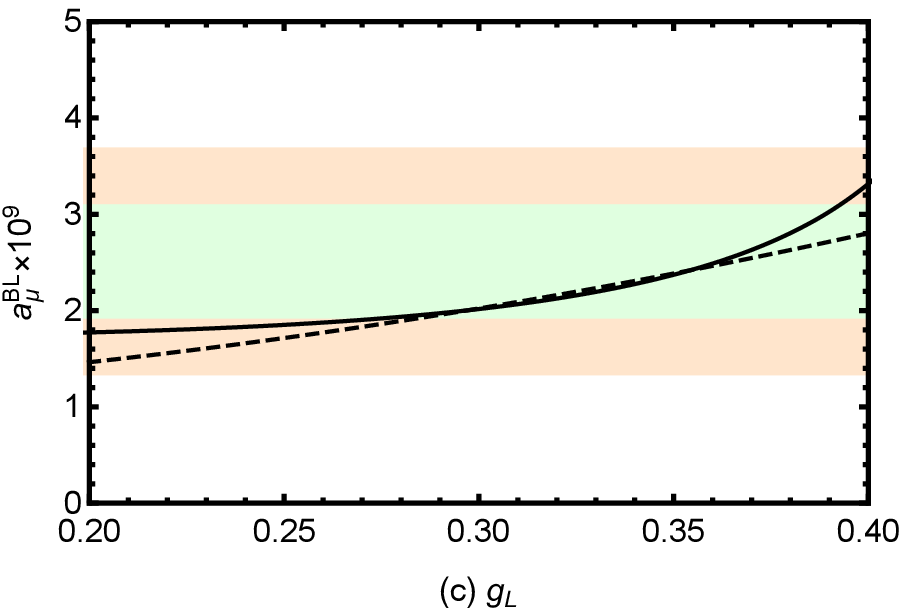}
\caption{The BLMSSM contributions to muon MDM ($a^{BL}_{\mu}$) versus $m_1$(a), $m_2$(b) and $g_L$(c) are plotted by solid line (the mass eigenstate expressions) and dashed line (the mass insertion approximation expressions).}{\label {bj}}
\end{figure}

The comparisons of the results of the two expressions are shown in Fig.~\ref {bj}. The experimental limitations are denoted by the colored areas, where light green area represents 1$\sigma$, light orange area represents 2$\sigma$. The lines of these three figures are within the colored areas, which can well satisfy the experimental constraint. In Fig.~\ref {bj}(a), the two lines have similar behavior, that is, slowly increase and then slowly decrease. The two lines are relatively close, and there are two intersections. In Fig.~\ref {bj}(b), the two lines almost coincide in the area $m_2$ (1000,~1500)~{\rm GeV} and are very close in the remaining area. In Fig.~\ref {bj}(c), the two lines have an upward trend and almost coincide in the area $g_L$ (0.3,~0.35). Therefore, according to these diagrams, we can obtain that the results of the two expressions are very similar, and the accuracy of the MIA results is verified.

\subsection{The muon MDM by MIA}

In this subsection, we numerically calculate the BLMSSM contributions to muon MDM ($a^{BL}_{\mu}$). Based on the above analysis of the mass insertion approximation, $a^{BL}_{\mu}$ mainly depends on 9 parameters, i.e., $\tan{\beta}$,~$g_L$,~$m_1$,~$m_2$,~$m_L$,~$\mu_H$,~$m_{\tilde{\nu}_L}$,~$m_{{\tilde{\mu}}_R}$,
$m_{{\tilde{\mu}}_L}$. We take these parameters as free parameters and compute the BLMSSM contributions to the muon MDM for a given set of parameters, with the parameter $M_{SUSY}=1000 ~{\rm GeV}$.

\subsubsection{One-dimensional graphs}

In this part, we take
$m_1=300~{\rm GeV},~m_L=-300~{\rm GeV},~m_{\tilde{\nu}_L}=150~{\rm GeV}, m_{{\tilde{\mu}}_L}=700~{\rm GeV},~m_{{\tilde{\mu}}_R}=700~{\rm GeV}$, and plot the following $a^{BL}_{\mu}$ schematic diagram affected by different parameters.
The colored areas show the experimental limitations, with light green area for 1$\sigma$ and light orange area for 2$\sigma$.

\begin{figure}[ht]
\setlength{\unitlength}{5mm}
\centering
\includegraphics[width=5.0in]{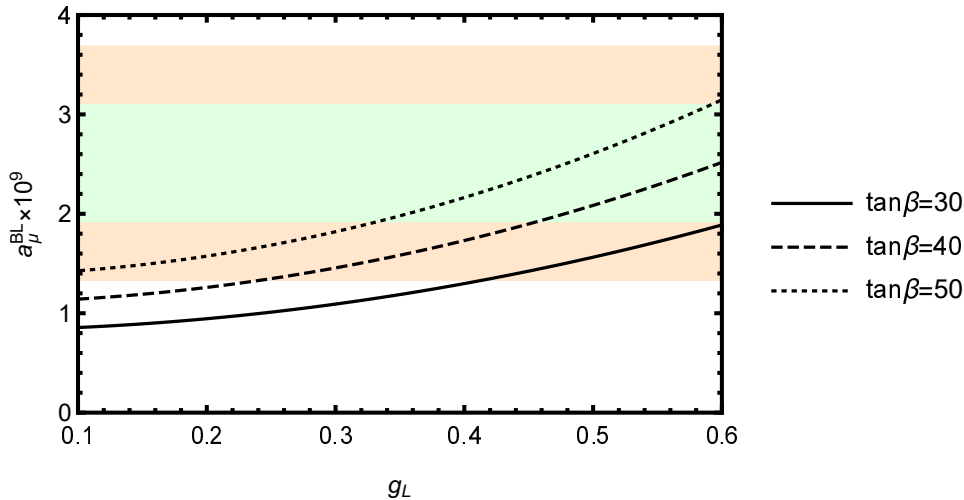}
\caption{The BLMSSM contributions to muon MDM ($a^{BL}_{\mu}$) versus $g_L$.}{\label {y1}}
\end{figure}

In Fig.~\ref {y1}, we plot the results versus $g_L$ with $m_2=1100~{\rm GeV}$ and $\mu_H=1100~{\rm GeV}$. Beyond MSSM, there is a parameter $g_L$ that corresponds to the coupling constant of the $U(1)_{L}$ gauge.
From the analysis by MIA, $g_L$ is an important parameter that appears in Eq.~(\ref{MIAe}).
It can be seen that from bottom to top are solid line ($\tan{\beta}=30$), dashed line ($\tan{\beta}=40$) and dotted line ($\tan{\beta}=50$), and the overall trend of the three lines is upward. This conclusion can be seen more intuitively from Eq.~(\ref{amumssm3}).
The dotted line is entirely in the colored areas, the dashed line part of $0.22-0.6$ is in the colored areas, the solid line part of $0.4-0.6$ is in the colored areas. That is to say, $\tan{\beta}$ is a sensitive parameter and larger $\tan{\beta}$ leads to larger $a^{BL}_{\mu}$.
The value of $a^{BL}_{\mu}$ is around $2.5\times10^{-9}$, and it can better meet the experimental limitations.

\begin{figure}[ht]
\setlength{\unitlength}{5mm}
\centering
\includegraphics[width=5.0in]{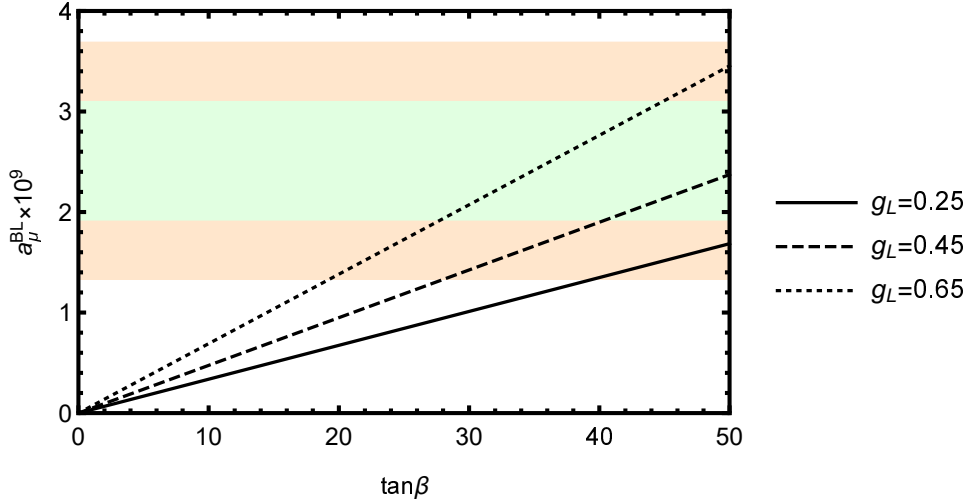}
\caption{The BLMSSM contributions to muon MDM ($a^{BL}_{\mu}$) versus $\tan\beta$.}{\label {y2}}
\end{figure}

Similarly, we take $m_2=1100~{\rm GeV}$ and $\mu_H=1100~{\rm GeV}$, and
plot the BLMSSM contributions to muon MDM varying with $\tan\beta$ in Fig.~\ref{y2}. The parameter $\tan\beta$ is ratio of the VEVs of the two Higgs doublets ($\tan\beta=\upsilon_{u} / \upsilon_{d}$). $\tan\beta$ is contained in each one-loop contribution and is proportional relationship.
The solid (dashed, dotted) line corresponds to the results with $g_L=0.25~(0.45, ~0.65)$. We find that $\tan\beta$ and $a^{BL}_{\mu}$ are positively correlated, which is the same effect reflected by Eqs.~(\ref{MIAa}),~(\ref{MIAb})$-$(\ref{MIAf}). The colored areas contain more parts of the dotted line. The characteristic obtained here is consistent with those obtained in Fig.~\ref {y1}. Large $g_L$ and $\tan{\beta}$ can lead to large BLMSSM contributions.
Thus, the contributions can be influenced obviously by the parameters $g_L$ and $\tan{\beta}$. The value of $a^{BL}_{\mu}$ is around $2.5\times10^{-9}$, which can well compensate for the deviation.

\begin{figure}
\setlength{\unitlength}{5mm}
\centering
\includegraphics[width=5.0in]{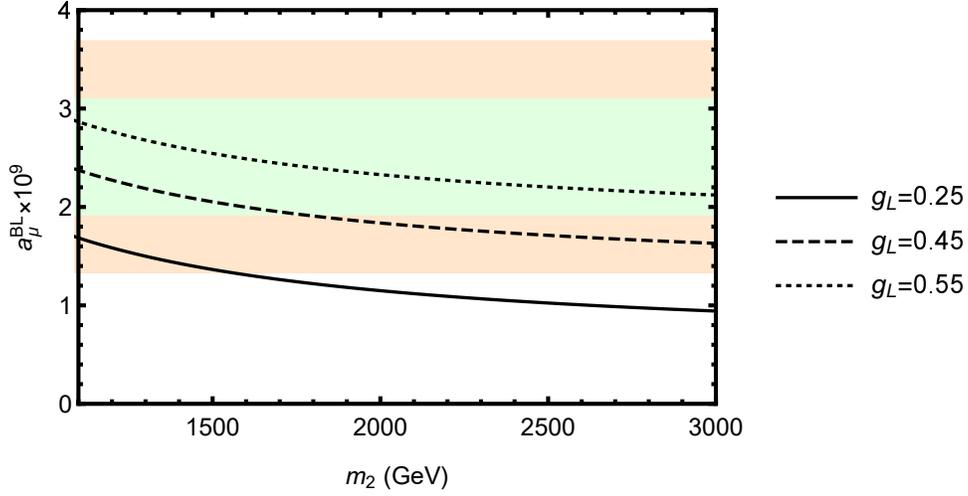}
\caption{The BLMSSM contributions to muon MDM ($a^{BL}_{\mu}$) versus $m_2$.}{\label {y3}}
\end{figure}

For better numerical results, we set $\tan{\beta}=50$ and $\mu_H=1100~{\rm GeV}$. The solid line ($g_L=0.25$), dashed line ($g_L=0.45$) and dotted line ($g_L=0.55$) varying with $m_2$ are shown in Fig.~\ref {y3}. $m_2$ expresses the particle mass of $\tilde{W}^0$ ($\tilde{W}^{\pm}$), which directly affects the one-loop contributions from Figs.~\ref{N2}(d),~\ref{N2}(f). The three lines are all decreasing functions, when $m_2$ turns large from 1100~{\rm GeV} to 3000~{\rm GeV}. The downward trend slowly becomes weak. The reason is that the contributions are proportional to $\sqrt{x_2}=\frac{m_2}{M_{SUSY}}$, but the effect of $m_2$ on the function is considerable and inversely proportional in Eqs.~(\ref{MIAd}),~(\ref{MIAf}) obtained by MIA. On the whole, the increase of $m_2$ leads to the slow decrease of $a^{BL}_{\mu}$. The dotted and dashed lines are all located in the colored areas. The dotted line can reach $2.9\times10^{-9}$, the dashed line can reach $2.4\times10^{-9}$ and the solid line can reach $1.7\times10^{-9}$.

\begin{figure}
\setlength{\unitlength}{5mm}
\centering
\includegraphics[width=5.0in]{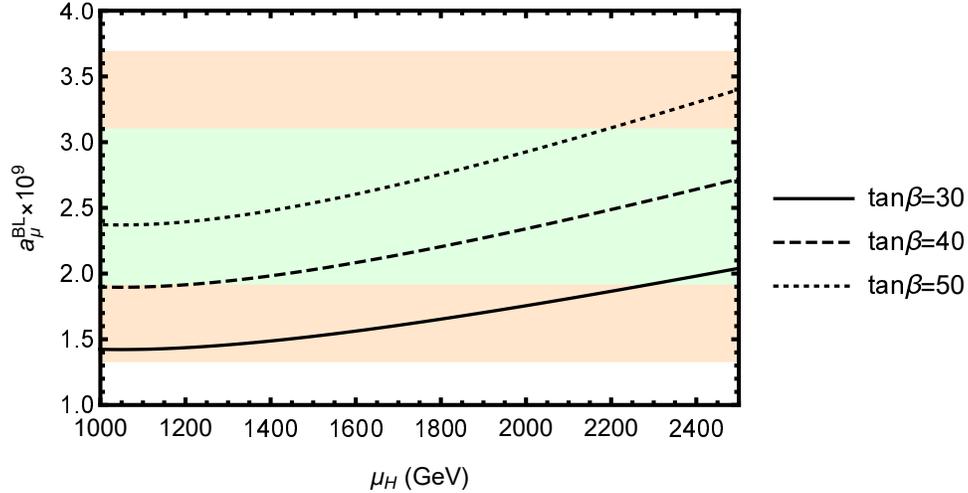}
\caption{The BLMSSM contributions to muon MDM ($a^{BL}_{\mu}$) versus $\mu_H$.}{\label {y4}}
\end{figure}

In addition, supposing the parameters with $g_L=0.45$ and $m_2=1100~{\rm GeV}$, we study the parameter $\mu_H$ influences on muon MDM in Fig.~\ref {y4}. $\mu_H$ is SUSY invariant Higgs mass, which exists in each contribution of Fig.~\ref{N2}. The solid line, dashed line and dotted line respectively correspond to the results with $\tan{\beta}=30$, $\tan{\beta}=40$ and $\tan{\beta}=50$. All three lines have a slight positive slope. Among the dominant terms, $a^{BL,(a)}_\mu$, $a^{BL,(e)}_\mu$, and $a^{BL,(f)}_\mu$ are proportional to $\sqrt{x_{\mu_H}}=\frac{m_{\mu_H}}{M_{SUSY}}$, but the function parts are inversely proportional to $\mu_H$ and have a relatively small effect. After combination, the effect of $\mu_H$ on muon MDM shows a slowly increasing relationship, when $\mu_H$ increases from 1000~{\rm GeV} to 2500~{\rm GeV}.
The three curves all are in the colored areas, which mean that $a^{BL}_{\mu}$ satisfies the experimental limitations under our assumption. The dotted line is at the top, that is, large $\tan{\beta}$ value results in larger $a^{BL}_{\mu}$. All three lines can exceed $2.0\times10^{-9}$.

\subsubsection{Multidimensional scatter plots graphs}

\begin{figure}[ht]
\setlength{\unitlength}{5mm}
\centering
\includegraphics[width=3.0in]{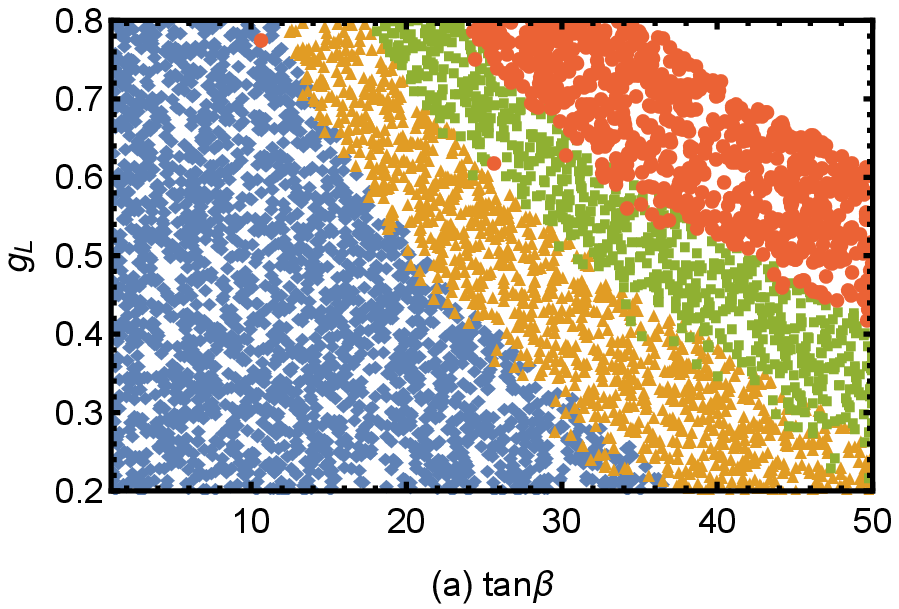}
\vspace{0.2cm}
\setlength{\unitlength}{5mm}
\centering
\includegraphics[width=3.15in]{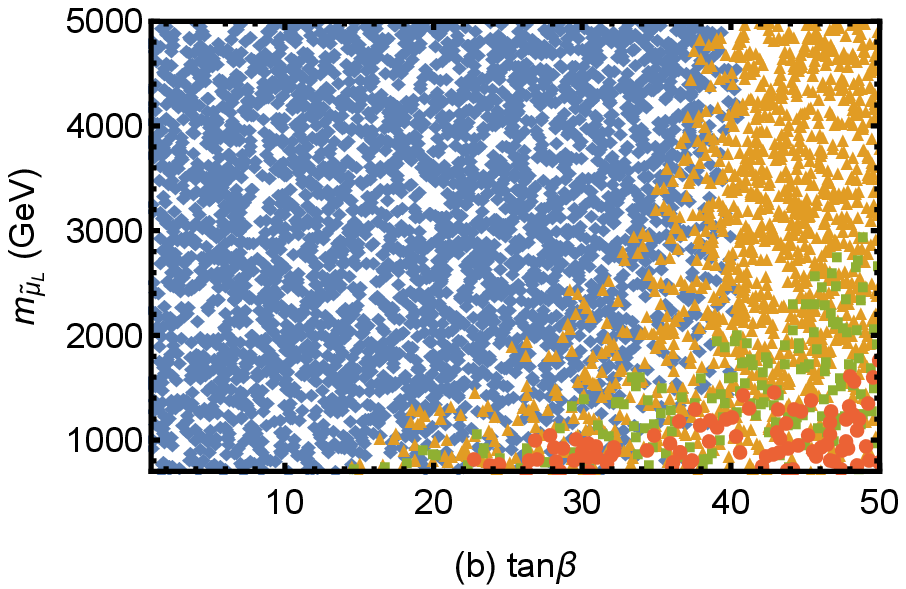}
\caption{$a^{BL}_{\mu}$ in $\tan{\beta}-g_L$ plane (a) and $\tan{\beta}-m_{\tilde{\mu}_L}$ plane (b).}{\label {D1}}
\end{figure}

\begin{figure}[ht]
\setlength{\unitlength}{5mm}
\centering
\includegraphics[width=3.15in]{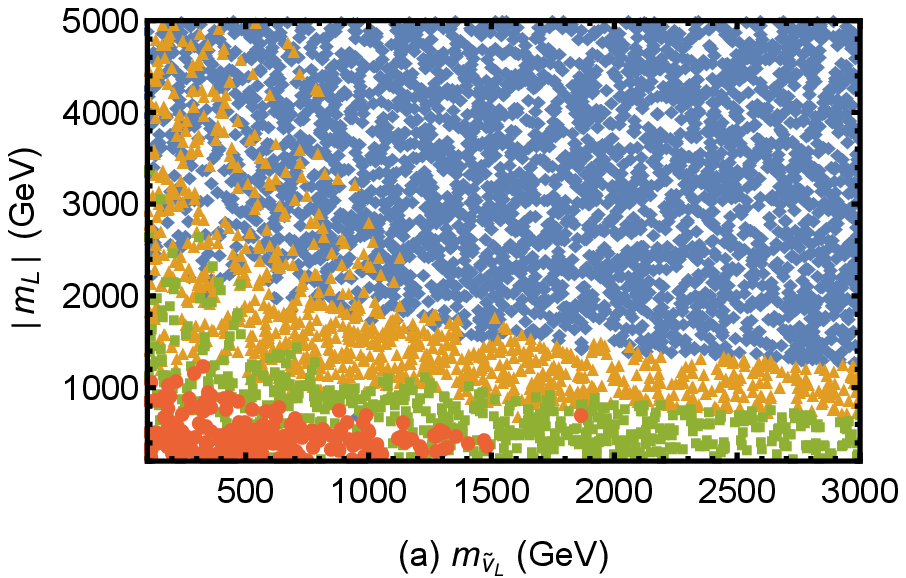}
\vspace{0.2cm}
\setlength{\unitlength}{5mm}
\centering
\includegraphics[width=3.0in]{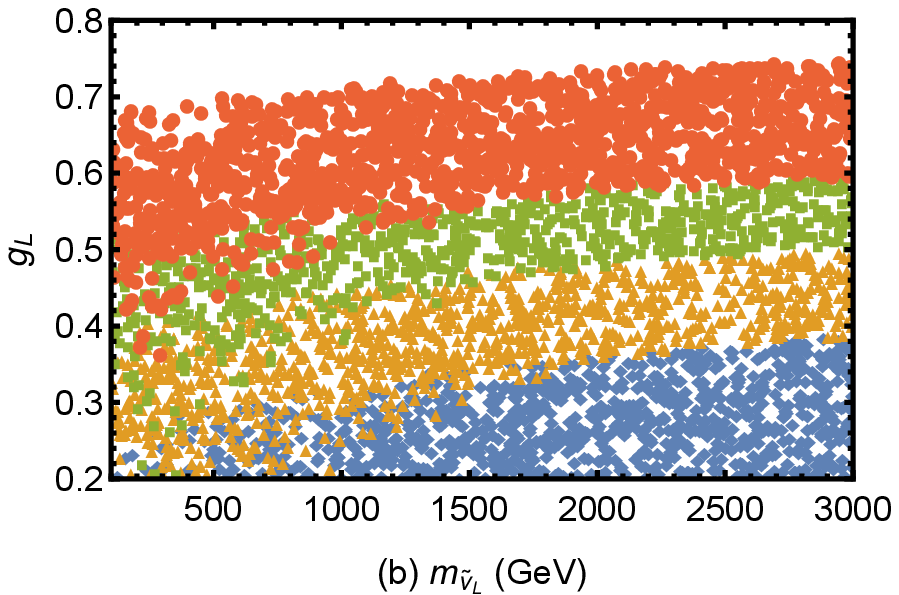}
\caption{$a^{BL}_{\mu}$ in $m_{\tilde{\nu}_L}-|m_L|$ plane (a) and $m_{\tilde{\nu}_L}-g_L$ plane (b).}{\label {D2}}
\end{figure}

\begin{figure}
\setlength{\unitlength}{5mm}
\centering
\includegraphics[width=3.1in]{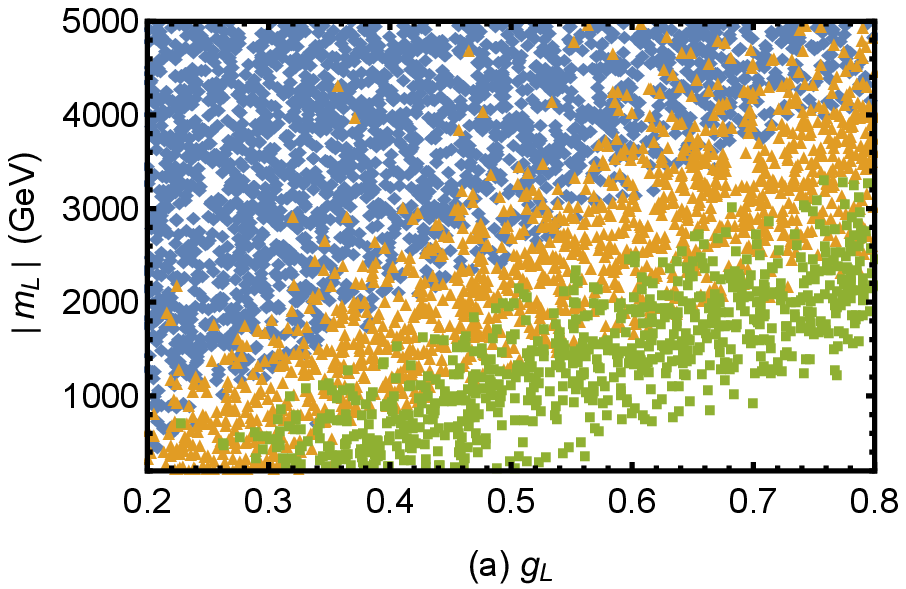}
\vspace{0.2cm}
\setlength{\unitlength}{5mm}
\centering
\includegraphics[width=3.1in]{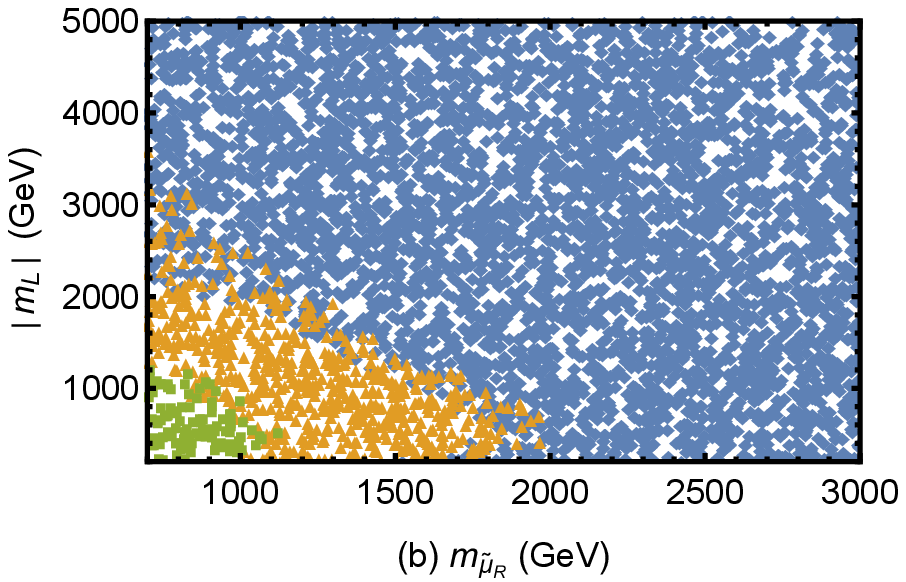}
\caption{$a^{BL}_{\mu}$ in $g_L-|m_L|$ plane (a) and $m_{\tilde{\mu}_R}-|m_L|$ plane (b).}{\label {D3}}
\end{figure}

In this part, we carry out numerical analysis by scanning free parameters and explore the region to explain the BLMSSM contributions to muon MDM. The random ranges of input parameters are as follows:
\begin{eqnarray}
&&\tan{\beta} \supset [1,50],~~~g_L \supset [0.2,0.8],~~~m_1 \supset [100,3000]~{\rm GeV},\nonumber\\&&m_2 \supset [1000,3000]~{\rm GeV},~~\mu_H \supset [1000,3000]~{\rm GeV},~~m_{\tilde{\nu}_L} \supset [100,3000]~{\rm GeV},\nonumber\\&&m_{{\tilde{\mu}}_L} \supset [700,3000]~{\rm GeV},~~m_{{\tilde{\mu}}_R} \supset [700,3000]~{\rm GeV},~~|m_L| \supset [200,5000]~{\rm GeV}.
\end{eqnarray}
In Table \ref {t3}, we show the markers for Figs.~\ref{D1}$-$\ref{D3}.

\begin{table}[ht]
\caption{ The meaning of shape style}
\begin{tabular}{|c|c|c|c|}
\hline
Shape style&Figs.~\ref{D1},\ref{D2}&Fig.~\ref{D3}\\
\hline
\textcolor{black-blue}{$\blacklozenge$} & $0<a^{BL}_{\mu}<10^{-9}$ & $0<a^{BL}_{\mu}<1.5 \times 10^{-9}$ \\
\hline
\textcolor{black-yellow}{$\blacktriangle$}&$10^{-9} \leqslant a^{BL}_{\mu}<1.5 \times 10^{-9}$ & $1.5 \times 10^{-9} \leqslant a^{BL}_{\mu}<2.0 \times 10^{-9}$ \\
\hline
\textcolor{black-green}{$\blacksquare$}&$1.5 \times 10^{-9} \leqslant a^{BL}_{\mu}<2.0 \times 10^{-9}$ & $2.0 \times 10^{-9} \leqslant a^{BL}_{\mu}<3.0 \times 10^{-9}$ \\
\hline
\textcolor{black-red}{$\bullet$}&$2.0 \times 10^{-9} \leqslant a^{BL}_{\mu}<3.0 \times 10^{-9}$ & $\backslash$ \\
\hline
\end{tabular}
\label{t3}
\end{table}

To better display sensitive parameters, we show the $a^{BL}_{\mu}$ in the $\tan{\beta}-g_L$ plane (a) and $\tan{\beta}-m_{\tilde{\mu}_L}$ plane (b) in Fig.~\ref{D1}. The bounds between \textcolor{black-blue}{$\blacklozenge$}, \textcolor{black-yellow}{$\blacktriangle$},
\textcolor{black-green}{$\blacksquare$} and \textcolor{black-red}{$\bullet$} are very obvious in Fig.~\ref{D1}(a). The blue part is displayed in a trapezoid and takes up a lot of space. The results represented by the remaining three colors show a slight radian. The red part is on the upper right corner, that is, large $\tan{\beta}$ and large $g_L$ can bring greater contributions. Similarly, \textcolor{black-blue}{$\blacklozenge$} also occupy a large number of positions in Fig.~\ref{D1}(b) and mainly in the wide area $1<\tan{\beta}<40$ and $700~{\rm GeV}<m_{\tilde{\mu}_L}<5000~{\rm GeV}$. \textcolor{black-red}{$\bullet$} concentrate in the narrow area
$\tan{\beta}~(22,50)$ and $m_{\tilde{\mu}_L}~(700,1600)~{\rm GeV}$. $m_{\tilde{\mu}_L}$ is left-handed smuon mass. The function parts of $a^{BL,(a)}_\mu$, $a^{BL,(c)}_\mu$, $a^{BL,(d)}_\mu$, and $a^{BL,(e)}_\mu$ are inversely proportional to $m_{\tilde{\mu}_L}$. Therefore, this means that light scalar muon improves the BLMSSM contributions to muon MDM.

We plot $a^{BL}_{\mu}$ in the plane of $m_{\tilde{\nu}_L}$ versus $|m_L|$ by the left diagram in the Fig.~\ref{D2}, and the right diagram shows the relation between $a^{BL}_{\mu}$, $m_{\tilde{\nu}_L}$ and $g_L$. One can find that the styles of Fig.~\ref{D2}(a) and Fig.~\ref{D1}(b) are similar. In Fig.~\ref{D2}(a), the blue area is the most.
In the range $800~{\rm GeV}< |m_L| <1200~{\rm GeV}$, \textcolor{black-yellow}{$\blacktriangle$} occupy much space. \textcolor{black-green}{$\blacksquare$} concentrate in the narrow area
$|m_L|<800~{\rm GeV}$ and $1500 ~{\rm GeV}<m_{\tilde{\nu}_L}<3000 ~{\rm GeV}$. \textcolor{black-red}{$\bullet$} denote large contributions to $a^{BL}_{\mu}$ that are
concentrate in the area $|m_L| <1000~{\rm GeV}$ and $100 ~{\rm GeV}<m_{\tilde{\nu}_L}<1500 ~{\rm GeV}$. $m_L$ expresses the mass of new gaugino $\lambda_{L}$ beyond MSSM. We take $m_L$ as a negative value in the Eq.~(\ref{MIAe}) and can easily find that this contribution is proportional to $\sqrt{x_L}=\frac{m_L}{M_{SUSY}}$. For more convenience, we take $|m_L|$ as the ordinate. These indicate that small $|m_L|$ and small $m_{\tilde{\nu}_L}$ can lead to large corrections. In Fig.~\ref{D2}(b), the layers are distinct, with \textcolor{black-yellow}{$\blacktriangle$},
\textcolor{black-green}{$\blacksquare$} and \textcolor{black-red}{$\bullet$}  arched. In the case of $m_{\tilde{\nu}_L}=3000~{\rm GeV}$, we can find these laws. When $g_L < 0.38$, the space is filled with \textcolor{black-blue}{$\blacklozenge$}. The red, green and blue parts correspond to $0.38<g_L < 0.5$, $0.5<g_L < 0.6$ and $0.6<g_L < 0.74$, respectively. $m_{\tilde{\nu}_L}$ denotes the mass of the left-handed neutrino, which causes a change in $a^{BL,(f)}_\mu$ by directly affecting the function part of Eq.~(\ref{MIAf}). The final effect is that $a^{BL}_\mu$ is inversely proportional to $m_{\tilde{\nu}_L}$. In the whole, small $m_{\tilde{\nu}_L}$ and large $g_{L}$ can obviously improve the corrections to $a^{BL}_\mu$.

Furthermore, we show the relationship between $g_L$ and $|m_L|$,  $m_{\tilde{\mu}_R}$ and $|m_L|$ in Fig.~\ref{D3}. It is worth noting that the meanings of \textcolor{black-blue}{$\blacklozenge$}, \textcolor{black-yellow}{$\blacktriangle$} and \textcolor{black-green}{$\blacksquare$} in Fig.~\ref{D3} are inconsistent with those in  Figs.~\ref{D1},\ref{D2}. The specific meanings are shown in right part of Table~\ref{t3}. The left figure and the right figure have obvious stratification and strong regularity. In Fig.~\ref{D3}(a), the results are divided into two parts by the diagonal. \textcolor{black-blue}{$\blacklozenge$} concentrate in the upper left of the diagonal and \textcolor{black-yellow}{$\blacktriangle$} and \textcolor{black-green}{$\blacksquare$} mainly distribute at the bottom right of the diagonal. In Fig.~\ref{D3}(b), the whole space is covered. \textcolor{black-green}{$\blacksquare$} concentrate in the narrow area
$m_{\tilde{\mu}_R}~(700,1000)~{\rm GeV}$ and $|m_L|~(200,1000)~{\rm GeV}$.  \textcolor{black-yellow}{$\blacktriangle$} occupy much space in the range $700~{\rm GeV}< |m_L| <2000~{\rm GeV}$ and $1000~{\rm GeV}< m_{\tilde{\mu}_R} <1800~{\rm GeV}$. The blue part occupies all the remaining positions. $m_{\tilde{\mu}_R}$ is right-handed smuon mass. Only the function parts of $a^{BL,(a)}_\mu$, $a^{BL,(b)}_\mu$ and $a^{BL,(e)}_\mu$ contain $m_{\tilde{\mu}_R}$ and are inversely proportional to $m_{\tilde{\mu}_R}$. The results imply that large $m_{\tilde{\mu}_R}$ and large $|m_L|$ can diminish the BLMSSM contributions to muon MDM. Based on the above description, we can be more clear about the contribution of the above parameters.

\subsection{The lepton EDM by MIA}

In this subsection, we research and analyze the one-loop contributions of the lepton
$(e, \mu, \tau)$ EDM in the frame work of CP violating BLMSSM via the mass insertion approximation. According to Part B of Section III, $d^{BL}_l$ mainly depends on 15 parameters, i.e., $\tan{\beta}$,~$g_L$,~$M_1$,~$M_2$,~$M_L$,~$M_ {\mu_H}$,~$M_{\mu_L}$,~$m_{\tilde{\nu}_L}$,~$m_{{\tilde{L}}_R}$,
$m_{{\tilde{L}}_L}$,~$\theta_1$,~$\theta_2$,~$\theta_{\mu_H}$,~$\theta_{\mu_L}$ and $\theta_{L}$. We take these parameters as free parameters and calculate the BLMSSM contributions to the lepton
$(e, \mu, \tau)$ EDM for a given set of parameters, and fix the parameter $M_{SUSY}=1000 ~{\rm GeV}$.

\subsubsection{The electron EDM}

\begin{figure}[ht]
\setlength{\unitlength}{5mm}
\centering
\includegraphics[width=5.0in]{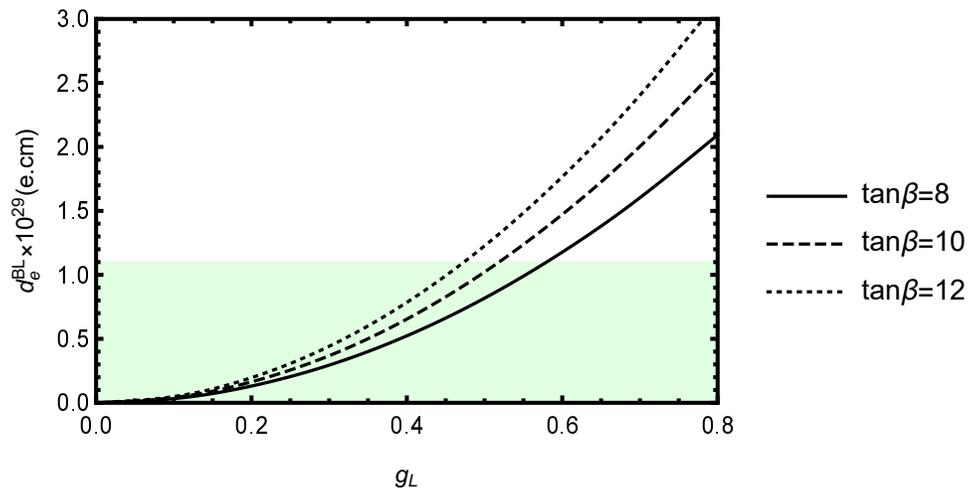}
\caption{With $\theta_{\mu_L}=\pi/1000$, and $ \theta_1=\theta_2=\theta_{\mu_H}=\theta_L=0$, the BLMSSM contributions to electron EDM ($d^{BL}_e$) versus $g_L$.}{\label {1b}}
\end{figure}

\begin{figure}[ht]
\setlength{\unitlength}{5mm}
\centering
\includegraphics[width=5.0in]{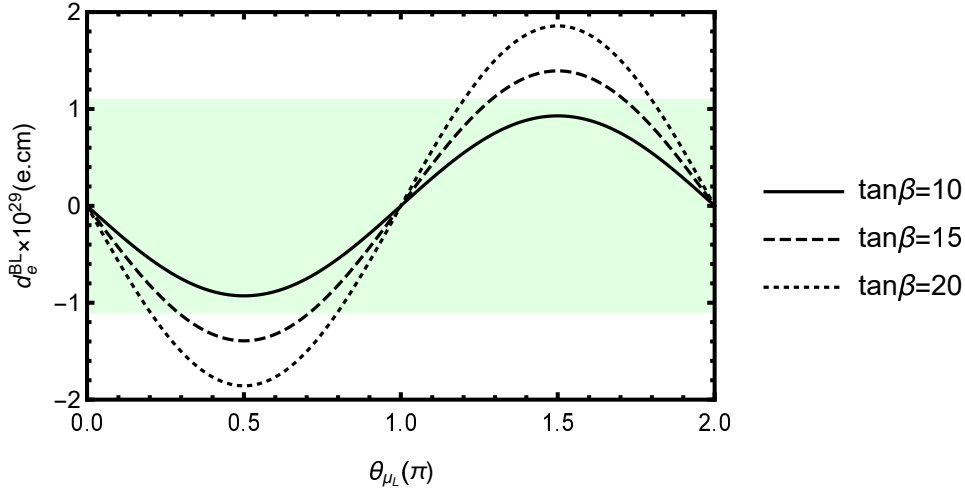}
\caption{With~$\theta_1=\theta_2=\theta_{\mu_H}=\theta_L=0$, the BLMSSM contributions to electron EDM ($d^{BL}_e$) versus $\theta_{\mu_L}$.}{\label {1a}}
\end{figure}

\begin{figure}[ht]
\setlength{\unitlength}{5mm}
\centering
\includegraphics[width=5.0in]{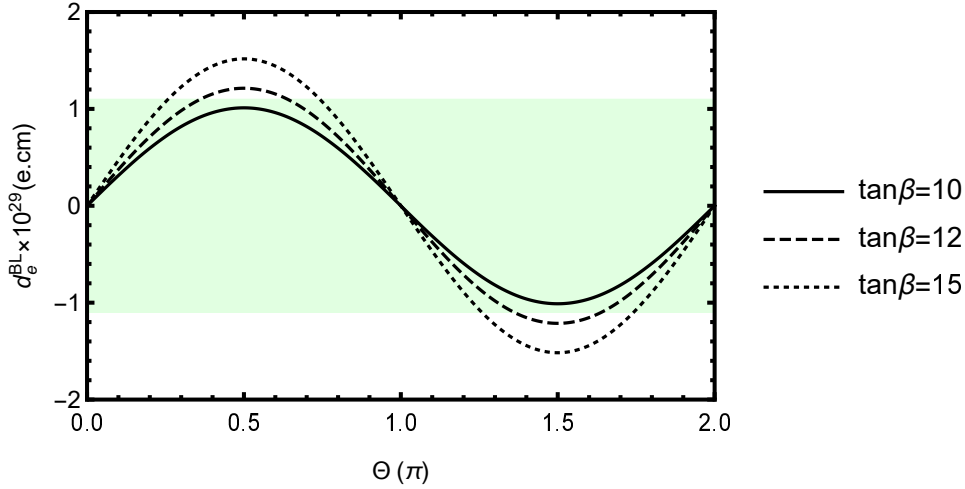}
\caption{With $ \theta_1=\theta_2=\theta_L=0$, the BLMSSM contributions to electron EDM ($d^{BL}_e$) versus $\Theta$ ($\Theta=\theta_{\mu_L}=\theta_{\mu_H}$).}{\label {1c}}
\end{figure}

First of all, we discuss the electron EDM because its experimental upper limit is very strict. The analysis consists of three parts: 1. make the CP violating phases small $O(10^{-2}-10^{-3})$; 2. increase the particle mass to the several 10 TeV range; 3. make internal cancellations between phases.
The light green areas of Figs.~\ref{1b},\ref{1a},\ref{1c} represent the experimental limitations of electron EDM.

a. Small phases

Supposing $M_1=800~{\rm GeV}$,~$M_2=1100~{\rm GeV}$,~$M_L=-3000~{\rm GeV}$,~$M_ {\mu_H}=1100~{\rm GeV}$,~$M_ {\mu_L}=1100~{\rm GeV}$,~$m_{\tilde{\nu}_L}=300~{\rm GeV}$,~$m_{{\tilde{L}}_R}=1800~{\rm GeV}$,
~$m_{{\tilde{L}}_L}=1800~{\rm GeV}$,~$\theta_1=\theta_2=\theta_{\mu_H}=\theta_L=0$ and $\theta_{\mu_L}=\pi/1000$, we study the contributions from $g_L$ to electron EDM. We plot the solid line,
dashed line and dotted line versus $g_L (0 - 0.8)$ corresponding to $\tan{\beta}=
(8,10,12)$ in Fig.~\ref{1b}. Obviously, these three lines are all increasing functions of $g_L$. With $\tan{\beta}=12$,~$\tan{\beta}=10$ and $\tan{\beta}=8$, $d^{BL}_e$ can satisfy the experimental bound in the $g_L$ region $(0 - 0.45)$,~$(0 - 0.5)$ and $(0 - 0.6)$, respectively. Therefore, a small phase and particle mass in the TeV range can satisfy the experimental constraints of electron EDM. However, a small phase represents fine tuning unless it occurs naturally, for example as a loop correction.

b. Mass suppression

The parameters $g_L=0.1$,~$M_1=800~{\rm GeV}$,~$M_2=1100~{\rm GeV}$,~$M_L=4500~{\rm GeV}$,~$M_ {\mu_H}=1100~{\rm GeV}$,~$m_{\tilde{\nu}_L}=300~{\rm GeV}$, and $\theta_1=\theta_2=\theta_{\mu_H}=\theta_L=0$ are adopted.
The remaining parameters are large masses of particles, i.e., $M_ {\mu_L}=30000~{\rm GeV}$,~$m_{{\tilde{L}}_R}=40000~{\rm GeV}$ and $m_{{\tilde{L}}_L}=40000~{\rm GeV}$. With $\tan{\beta}=(10,15,20)$, the results are shown by the solid line, dashed line and dotted line respectively in Fig.~\ref{1a}. $\theta_{\mu_L}$ is the CP violating phase of new parameter $\mu_L$. $\mu_L$ relates with sneutrino mass squared matrix and lepton neutralino mass matrix. The shapes of the three lines are consistent, and they are very similar as $-\sin\theta_{\mu_L}$. For $\theta_{\mu_L}$ from $\pi$ to $2\pi$, the dotted line is up the dashed line and the dashed line is up the solid line. Total solid line and most parts of dashed and dotted lines are in the light green area. That is to say, particle mass in the several 10 TeV range and a normal phase can easily satisfy the experimental limitations of electron EDM. However, this violates the naturalness. Obviously, the several 10 TeV range even may be out of reach of LHC.

\newpage

c. Internal cancellations

To sum up, we discuss the third possible method, which is internal cancellations among the different phases of the electron EDM. We study the numerical results versus $\Theta$ ($\Theta=\theta_{\mu_L}=\theta_{\mu_H}$) with
$g_L=0.5$,~$M_1=800~{\rm GeV}$,~$M_2=2200~{\rm GeV}$,~$M_L=1000~{\rm GeV}$,~$M_ {\mu_H}=1800~{\rm GeV}$,~$M_ {\mu_L}=1200~{\rm GeV}$,~$m_{\tilde{\nu}_L}=1200~{\rm GeV}$,~$m_{{\tilde{L}}_R}=700~{\rm GeV}$,
~$m_{{\tilde{L}}_L}=700~{\rm GeV}$ and $\theta_1=\theta_2=\theta_L=0$. The corresponding results
are plotted by the solid line, dashed line and dotted line in Fig.~\ref{1c}. The three lines all look like $\sin\Theta$. At these points $\Theta=0$,~$\Theta=\pi$, and $\Theta=2\pi$,
there is none CP violating effect and $d^{BL}_e=0$ is reasonable. The largest values of the three lines are respectively $1.01\times10^{-29}$ e.cm, $1.21\times10^{-29}$ e.cm and $1.51\times10^{-29}$ e.cm. With $\tan{\beta}=10$, the solid line all in the experiment constraint. This possible method has normal size CP violating phases $O(1)$ and particle mass in the TeV range. The obtained results can also more easily satisfy the experimental limitations of electron EDM.

\subsubsection{The muon EDM}

\begin{figure}[ht]
\setlength{\unitlength}{5mm}
\centering
\includegraphics[width=5.0in]{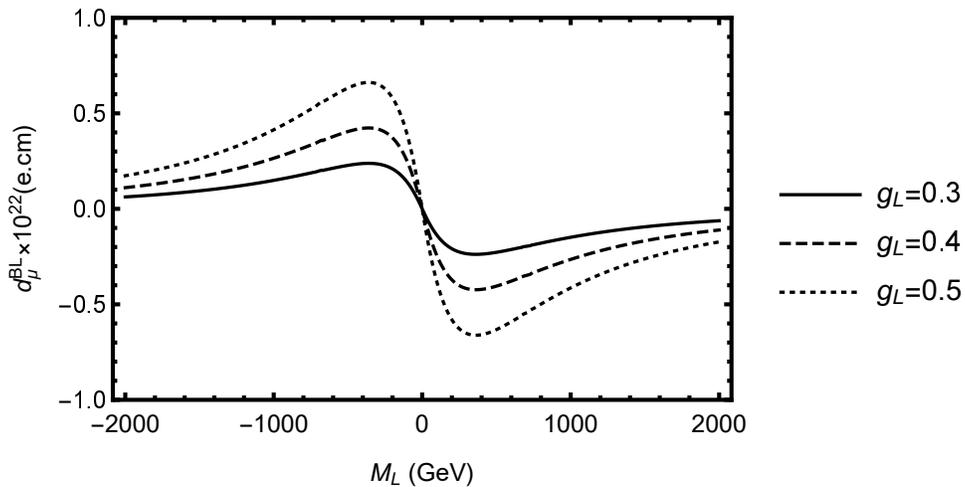}
\caption{With $\theta_L=\pi/3$, and $ \theta_1=\theta_2=\theta_{\mu_H}=\theta_{\mu_L}=0$, the BLMSSM contributions to muon EDM ($d^{BL}_\mu$) versus $M_L$.}{\label {21}}
\end{figure}

At present, the experimental upper bound of muon EDM is $|d^{exp}_\mu| < 1.8\times10^{-19}$ e.cm. In the part, we adopt the parameters as $\tan{\beta}=15$,~$M_1=800~{\rm GeV}$,~$M_2=1100~{\rm GeV}$,~$M_ {\mu_H}=1100~{\rm GeV}$,~$M_ {\mu_L}=2500~{\rm GeV}$,~$m_{\tilde{\nu}_L}=150~{\rm GeV}$,~$m_{{\tilde{L}}_R}=700~{\rm GeV}$,
$m_{{\tilde{L}}_L}=700~{\rm GeV}$, $\theta_1=\theta_2=\theta_{\mu_H}=\theta_{\mu_L}=0$ and $\theta_L=\pi/3$. We study $d^{BL}_\mu$ versus $M_L$ with $g_L=(0.3,0.4,0.5)$, and the results are plotted by the solid line, dashed line and dotted line in Fig.~\ref{21}. $M_L$ is the gaugino mass for the new gaugino $\lambda_L$. In Eq.~(\ref{edmjiexi}), $d^{BL,(e)}_l$ is proportional to $\sqrt{x_L}=\frac{|M_L|}{M_{SUSY}}$. The dashed line reaches $2.38\times10^{-23}$ e.cm as $M_L=-360~{\rm GeV}$. When $|M_L|>360~{\rm GeV}$, the absolute values of the numerical results shrink with the enlarging $|M_L|$. Larger $g_L$ results in larger $d^{BL}_\mu$, when the other parameters are same. All numerical results are around the order of  $10^{-22}$ e.cm, which is almost three-order smaller than muon EDM upper bound.

\subsubsection{The tau EDM}

\begin{figure}[ht]
\setlength{\unitlength}{5mm}
\centering
\includegraphics[width=5.0in]{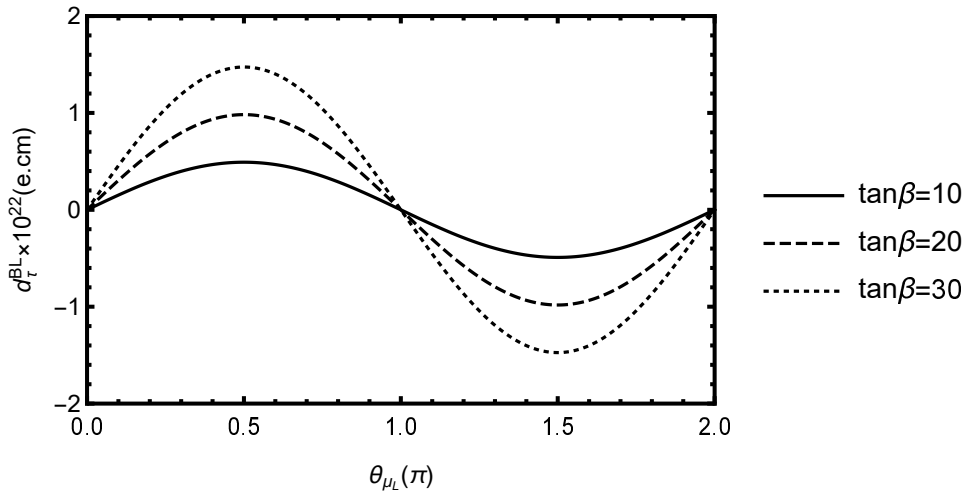}
\caption{With $\theta_1=\theta_2=\theta_{\mu_H}=\theta_L=0$, the BLMSSM contributions to tau EDM ($d^{BL}_\tau$) versus $\theta_{\mu_L}$.}{\label {31}}
\end{figure}

Among the lepton EDM bounds, the tau EDM has the loosest experimental upper limit, which is about $10^{-17}$ e.cm. Now, supposing $g_L=0.6$,~$M_1=800~{\rm GeV}$,~$M_2=1100~{\rm GeV}$,~$M_L=-500~{\rm GeV}$,~$M_ {\mu_H}=1100~{\rm GeV}$,~$M_ {\mu_L}=1100~{\rm GeV}$,~$m_{\tilde{\nu}_L}=150~{\rm GeV}$,~$m_{{\tilde{L}}_R}=1500~{\rm GeV}$,
$m_{{\tilde{L}}_L}=1500~{\rm GeV}$, and $\theta_1=\theta_2=\theta_{\mu_H}=\theta_L=0$, we study the influence
of $\theta_{\mu_L}$ on tau EDM ($d^{BL}_\tau$). In Fig.~\ref{31}, the solid line, dashed line and dotted line respectively correspond to $\tan{\beta}=(10,20,30)$ and their numerical results are all within the experimental limits of tau EDM.
There is none CP violating effect and $d^{BL}_\tau=0$ is reasonable, as $\theta_{\mu_L}=0$,~$\theta_{\mu_L}=\pi$ and $\theta_{\mu_L}=2\pi$. When $\theta_{\mu_L}=0.5\pi$ and $\theta_{\mu_L}=1.5\pi$, the absolute value of $d^{BL}_\tau$ reaches the maximum. The dotted line can arrive at $1.47\times10^{-22}$ e.cm. Generally speaking, the results are about five-order smaller than the experimental bound of the tau EDM.

\section{Discussion and conclusion}

In the framework of the BLMSSM, we study the one-loop contributions to the
muon MDM and the lepton $(e, \mu, \tau)$ EDM. During the analysis, the mass insertion approximation is used to more clearly display sensitive parameters. All parameters used can satisfy the latest experimental data. The relative error ($ME-MIA \over ME$) between the mass eigenstate (ME) and the mass insertion approximation (MIA) is about $1\%$, which is shown as a narrow area in the figure. Therefore, the accuracy of the MIA expressions is verified.

As we mentioned before, there are dominant three parts on which $a^{BL}_{\mu}$ depends, i.e., $a^{BL,(a)}_\mu$, $a^{BL,(e)}_\mu$ and $a^{BL,(f)}_\mu$. We take $\tan{\beta}$,~$g_L$,~$m_1$,~$m_2$,~$m_L$,~$\mu_H$,~$m_{\tilde{\nu}_L}$,~$m_{{\tilde{\mu}}_R}$ and $m_{{\tilde{\mu}}_L}$ as free parameters. Among them,
$\tan{\beta}$,~$g_L$,~$m_L$ and $\mu_H$ are more sensitive parameters.
$a^{BL}_{\mu}$ is an increasing function of $\tan\beta,~g_L,~\mu_H$ and decreasing
function of $m_2$. Small $|m_L|$ and small $m_{\tilde{\nu}}$ can improve the BLMSSM contributions to muon MDM. In our used parameter space, the contributions to muon MDM can easily reach its upper bound and even exceed it. Our best numerical result of $a^{BL}_{\mu}$ is around $2.5 \times 10^{-9}$, which can well compensate the departure between the experiment data and the SM prediction.

The effects of the CP violating phases $\theta_1,\theta_2,\theta_{\mu_H},\theta_{\mu_L},\theta_L$ on the lepton $(e, \mu, \tau)$ EDM are researched. Using MIA, the impact of the CP violating phases on $d^{BL}_l$ can be observed more intuitively.
The addition of two new CP violating sources ($\theta_{\mu_L}$,$\theta_L$) and the coupling constant $g_L$ can enhance the one-loop contributions from the neutralino-slepton diagram by several orders. The upper bound of electron EDM is $1.1\times10^{-29}$ e.cm, which is the most stringent. This poses a challenge to the BLMSSM parameter space.
By using the method of two phases cancellation, the contributions to electron EDM can be controlled below the experimental limit in our parameter space. The numerical results of muon EDM and tau EDM are several orders smaller than their upper limits, and at the order of  $10^{-22}$ e.cm in our used parameter space. With the improvement of technical accuracy, the lepton EDM may be detected by the experiments in the near future.

\begin{acknowledgments}

This work is supported by National Natural Science Foundation of China (NNSFC)
(No.12075074), Natural Science Foundation of Hebei Province
(A2020201002, A202201022, A2022201017), Natural Science Foundation of Hebei Education Department (QN2022173), Post-graduate's Innovation Fund Project of Hebei University (HBU2022ss028, HBU2023SS043), the youth top-notch talent support program of the Hebei Province.
\end{acknowledgments}


\newpage




\begin{thebibliography}{50}
\vspace{3mm}

\bibitem{g2rep2020}T. Aoyama, N. Asmussen, M. Benayoun, et al., \emph{Phys. Rept.} {\bf 887} (2020) 1.
\bibitem{GWB}G.W. Bennett, et al., \emph{Phys. Rev. D} {\bf73} (2006) 072003.
\bibitem{AKDN1}A. Keshavarzi, D. Nomura, T. Teubner, \emph{Phys. Rev. D} {\bf97} (2018) 114025.
\bibitem{GCMH}G. Colangelo, M. Hoferichter,  P. Stoffer, \emph{J. High Energy Phys.} {\bf 02} (2019) 006.
\bibitem{MHBL}M. Hoferichter, B.L. Hoid, B. Kubis, \emph{J. High Energy Phys.} {\bf 08} (2019) 137.
\bibitem{MDAH}M. Davier, A. Hoecker, B. Malaescu, et al., \emph{Eur. Phys. J. C.}  {\bf80} (2020) 241.
\bibitem{AKDN2}A. Keshavarzi, D. Nomura, T. Teubner, \emph{Phys. Rev. D} {\bf 101} (2020) 014029.
\bibitem{TBPA}T. Blum, P.A. Boyle, V. Gulpers, et al., \emph{Phys. Rev. Lett.} {\bf121} (2018) 022003.
\bibitem{TAMH}T. Aoyama, M. Hayakawa, T. Kinoshita, et al., \emph{Phys. Rev. Lett.} {\bf 109} (2012) 111808.
\bibitem{GCFH}G. Colangelo, F. Hagelstein, M. Hoferichter, et al., \emph{J. High Energy Phys.}  {\bf03} (2020) 101.
\bibitem{GECS}G. Eichmann, C.S. Fischer, R. Williams, \emph{Phys. Rev. D} {\bf 101} (2020) 054015.
\bibitem{TBNC}T. Blum, N. Christ, M. Hayakawa, et al., \emph{Phys. Rev. Lett.} {\bf124} (2020) 132002.
\bibitem{TATK}T. Aoyama, T. Kinoshita, M. Nio, \emph{Atoms} {\bf 7} (2019) 28.
\bibitem{ACWJ}A. Czarnecki, W.J. Marciano, A. Vainshtein, \emph{Phys. Rev. D} {\bf67} (2003) 073006.
\bibitem{CGDS}C. Gnendiger, D. Stockinger, H.S. Kim, \emph{Phys. Rev. D} {\bf 88} (2013) 053005.
\bibitem{had2}M.T. Hansen, A. Patella, \emph{J. High Energy Phys.} {\bf 10} (2020) 029.
\bibitem{muon2}H. Davoudiasl, W.J. Marciano, \emph{Phys. Rev. D} {\bf98} (2018) 075011.
\bibitem{mdm2} K. Hagiwara, A. Keshavarzi, A.D. Martin, et al., \emph{Nucl. Part. Phys. Proc.} {\bf287-288} (2017) 33-38.
\bibitem{046}Muon g-2 Collaboration, \emph{Phys. Rev. Lett.} {\bf126} (2021) 141801.


\bibitem{EDM1}A. Shindler, \emph{Eur. Phys. J. A.} {\bf57} (2021) 128 .
\bibitem{de} J. Baron, et al., (ACME Collaboration) \emph{Science} {\bf343} (2014) 269.
\bibitem{de1} V. Andreev, et al., (ACME Collaboration) \emph{Nature} {\bf562} (2018) 355-60.
\bibitem{de2} A. Crivellin, M. Hoferichter, P.S. Wellenburg, \emph{Phys. Rev. D} {\bf98} (2018) 113002.
\bibitem{pdg2022}R.L. Workman, et al., (Particle Data Group), \emph{Prog. Theor. Exp. Phys. } {\bf2022} (2022) 083C01.
\bibitem{deSM1}M.E. Pospelov and I.B. Khriplovich, \emph{Sov. J. Nucl. Phys.} {\bf 53} (1991) 638.
\bibitem{deSM2}M. Pospelov and A. Ritz, \emph{Phys. Rev. D.} {\bf 89} (2014) 056006.



\bibitem{susy1}S. Heinemeyer, D. St\"{o}ckinger, G. Weiglein, \emph{Nucl. Phys. B} {\bf690} (2004) 62.
\bibitem{susy2}S. Heinemeyer, D. St\"{o}ckinger, G. Weiglein, \emph{Nucl. Phys. B} {\bf699} (2004) 103.

\bibitem{wx3}F. Wang, L. Wu, Y. Xiao, et al., \emph{Nucl. Phys. B.} {\bf 970} (2021) 115486 [arXiv:2104.03262].
\bibitem{caojj}J.J. Cao, J.W. Lian, Y.S. Pan, et al.,  \emph{J. High Energy Phys.} {\bf09} (2021) 175 [arXiv:2104.03284].
\bibitem{caojj1}J.J. Cao, X.L. Jia, L. Meng, et al., [arXiv:2210.08769].
\bibitem{caojj2}J.J. Cao, J.W. Lian, Y.S. Pan, et al.,  \emph{J. High Energy Phys.} {\bf03} (2022) 203 [arXiv:2201.11490].
\bibitem{caojj3}J.J. Cao, F. Li, J.W. Lian, et al.,  \emph{Sci.China Phys.Mech.Astron.} {\bf65} (2022) 9, 291012 [arXiv:2204.04710].
\bibitem{YW1}W. Yin, \emph{J. High Energy Phys.} {\bf06} (2021) 029 [arXiv:2104.03259].
\bibitem{YW2}W. Yin, M. Yamaguchi, \emph{Phys.Rev.D} {\bf106} (2022) 3, 033007 [arXiv:2012.03928].

\bibitem{muon}S.M. Zhao, T.F. Feng, H.B. Zhang, et al., \emph{J. High Energy Phys.} {\bf11} (2014) 119 [arXiv: 1405.7561].
    \bibitem{o2}J.L. Yang, H.B. Zhang, C.X. Liu, et al., \emph{J. High Energy Phys.} {\bf08} (2021) 086.
\bibitem{o3}J.L. Yang, T.F. Feng, Y.L. Yan, et al., \emph{Phys. Rev. D.} {\bf99} (2019) 015002.
\bibitem{slh}L.H. Su, S.M. Zhao, X.X. Dong, et al., \emph{Eur. Phys. J. C} {\bf81} (2021) 433.
\bibitem{04}S.M. Zhao, L.H. Su, and X.X. Dong, et al., \emph{J. High Energy Phys.} {\bf03} (2022) 101.
\bibitem{o1}C.X. Liu, H.B. Zhang, J.L. Yang, et al., \emph{J. High Energy Phys.} {\bf04} (2020) 002.

\bibitem{NPdl1}F. del Aguila, M. B. Gavela, J. A. Grifols, et al., \emph{Phys. Lett. B} {\bf129} (1983) 473.
\bibitem{NPdl2}T. Ibrahim, P. Nath, \emph{Phys. Rev. D} {\bf60} (1999) 099902.
\bibitem{NPdl3}S. Atag, E. Gurkanli, \emph{J. High Energy Phys.} {\bf1606} (2016) 118.
\bibitem{NPdl4}T. Abe, N. Omoto, O. Seto, et al., \emph{Phys. Rev. D} {\bf98} (2018) 075029.
\bibitem{mssm}J. Rosiek, \emph{Phys. Rev. D} {\bf41} (1990) 3464.
\bibitem{mssm1}H.P. Nilles, \emph{Phys. Rept.} {\bf110} (1984) 1.
\bibitem{mssm2}H.E. Haber, G.L. Kane, \emph{Phys. Rept.} {\bf117} (1985) 75.
\bibitem{Z2015}S.M. Zhao, T.F. Feng, X.J. Zhan, et al., \emph{J. High Energy Phys.} {\bf07} (2015) 124.
\bibitem{EDM4}T. Ibrahim and P. Nath, \emph{Phys. Rev. D} {\bf 57} (1998) 478.
\bibitem{EDM5}T. Ibrahim and P. Nath, \emph{Phys. Rev. D} {\bf 58}(1998) 019901.


\bibitem{BLMSSM00}P.F. Perez and M.B. Wise, \emph{J. High Energy Phys.} {\bf08} (2011) 068.
\bibitem{BLMSSM000}P.F. Perez and M.B. Wise, \emph{Phys. Rev. D} {\bf82} (2010) 011901.
\bibitem{BLMSSM11}P.F. Perez, \emph{Phys. Rept.} {\bf597} (2015) 1-30.

\bibitem{BLMSSM1} P.F. Perez, \emph {Phys. Lett. B} {\bf711} (2012) 353 [arXiv:1201.1501].

\bibitem{BLMSSM2} J.M. Arnold, P.F. Perez, B. Fornal, et al., \emph {Phys. Rev. D} {\bf85} (2012) 115024 [arXiv:1204.4458].
\bibitem{weBLMSSM}T.F. Feng, S.M. Zhao, H.B. Zhang, et al., \emph { Nucl. Phys. B} {\bf 871} (2013) 223.

\bibitem{FM} E. Arganda, M.J. Herrero, R. Morales, et al., \emph{J. High Energy Phys.} {\bf03} (2016) 055.


\bibitem{lepton}T.F. Feng, L. Sun, X.Y. Yang, \emph {Nucl. Phys. B} {\bf800} (2008) 221-252.
\bibitem{dabeta1} T. Moroi, \emph {Phys. Rev. D} {\bf 53} (1996) 6565-6575.
\bibitem{dabeta2} D. Stockinger, \emph {J. Phys. G} {\bf34} (2007) R45-R92.
\bibitem{wx7}P. Athron, C. Balazs, D.H.J. Jacob, et al., \emph{J. High Energy Phys.} {\bf09} (2021) 080 [arXiv:2104.03691].
\bibitem{wx1}M. Endo, K. Hamaguchi, S. Iwamoto, et al., \emph{J. High Energy Phys.} {\bf 07} (2021) 075.
\bibitem{wx2}M. Chakraborti, L. Roszkowski and S. Trojanowski, \emph{J. High Energy Phys.} {\bf 05} (2021) 252 [arXiv:2104.04458].
\bibitem{wx4}P. Cox, C.C. Han, and T.T. Yanagida, \emph{Phys. Rev. D.} {\bf 104} (2021) 075035 [arXiv:2104.03290].
\bibitem{wx5}M.V. Beekveld, W. Beenakker, M. Schutten, et al., \emph{SciPost Phys.} {\bf11} (2021) 3, 049 [arXiv:2104.03245].
\bibitem{wx6}M. Chakraborti, S. Heinemeyer and I. Saha, \emph{Eur. Phys. J. C.} {\bf81} (2021) 12, 1114 [arXiv:2104.03287].
\bibitem{su1}CMS Collaboration, \emph{Phys. Lett. B} {\bf716} (2012) 30.
\bibitem{su2}ATLAS Collaboration, \emph{Phys. Lett. B} {\bf716} (2012) 1.
\bibitem{Zp5d1}ATLAS Collaboration, \emph{Phys. Lett. B} {\bf796} (2019) 68  [arXiv:1903.06248].





\end{thebibliography}
\end{document}